\documentclass[a4paper,fleqn,usenatbib]{mnras}
\usepackage{hyperref}
\usepackage{newtxtext,newtxmath}

\usepackage[T1]{fontenc}
\usepackage{ae,aecompl}

\usepackage{graphicx}	
\usepackage{amsmath}	
\usepackage{amssymb}	
\usepackage{multicol}

\title[Metallicity populations in dwarfs]{The distinct stellar metallicity populations of simulated Local Group dwarfs}

\author[A. Genina et al.]{\newauthor Anna Genina$^{1}$\thanks{E-mail: anna.genina@durham.ac.uk (AG)}, Carlos S. Frenk$^{1}$,
Alejandro Ben\'{i}tez-Llambay$^{1}$\thanks{E-mail: alejandro.b.llambay@durham.ac.uk (ABL)},
Shaun Cole$^{1}$, \newauthor Julio F. Navarro$^{2\thanks{Senior CIfAR Fellow (JFN)}}$,
Kyle A. Oman$^{3}$, Azadeh Fattahi$^{1}$
\\
$^{1}$ Institute for Computational Cosmology, Department of Physics, Durham University, South Road, Durham, DH1 3LE, UK\\
$^{2}$ Department of Physics \& Astronomy, University of Victoria, Victoria, BC, V8P 5C2, Canada\\
$^{3}$ Kapteyn Astronomical Institute, University of Groningen, Postbus 800, NL-9700 AV Groningen, The Netherlands}

\date{Accepted XXX. Received YYY; in original form ZZZ}

\pubyear{2018}

\begin{document}
\label{firstpage}
\pagerange{\pageref{firstpage}--\pageref{lastpage}}
\maketitle

\begin{abstract}
  A number of Local Group dwarf galaxies are known to have two
  spatially segregated stellar metallicity populations, a centrally
  concentrated metal-rich population and a more extended metal-poor
  population. In this work we discuss mechanisms that lead to the
  formation of two spatially segregated metallicity populations. Using a set of
  high-resolution hydrodynamical simulations of Local Group-like
  environments, we select a sample of satellite and field galaxies, spanning the
  stellar mass range $10^6-10^9$M$_{\odot}$, that exhibit bimodality in their metallicity distributions. Among those, we identify a subsample with a strong spatial segregation in the two populations. We
  find three distinct mechanisms for their formation. In field dwarfs and in a small fraction of satellites, a merger causes the metal-poor stars to migrate to larger radii and
  encourages the available gas to sink to the centre of the dwarf. Most of the gas is subsequently blown out of the halo through star
  formation feedback, but the remaining gas is consumed in the
  formation of a metal-rich population. In
  the exclusive case of satellites that have retained some of their gas at infall, it is the compression of
  this gas by ram pressure near pericentre that triggers the
  formation of metal-rich stars, whilst simultaneously preventing star formation at larger radii through stripping. Additionally, in a small number of
  field and satellite dwarfs, interactions with gaseous
  filaments and other galaxies can result in the formation of a metal-rich population. Regardless of the formation mechanism, a history of mergers typically enhances the spatial segregation. 
\end{abstract}


\section{Introduction}

Dwarf galaxies in the Local Group show a diversity of star formation
histories: some form the
majority of their stars in a short burst early on, while others
continue to form stars over their lifetime at varying rates \citep{dolphinSFH,sfrdwarfs2, sfrdwarfs,FastSlowGallart}. Some also exhibit signs of peculiar
kinematics and stellar substructure, suggesting a history of accretion
\citep{amoriscofornax,delpinofornax,delpinofornaxrotation,delpinoandromeda,battagliaSextansAccret}. A
number of satellites of the Milky Way and Andromeda, including
Sculptor, Fornax, Sextans, Andromeda~II and Carina display evidence of
distinct stellar metallicity and age populations
\citep{tolstoySculptor,battagliaSculptor,battagliaFornax,battagliaSextans,andromedaii,battagliaCarina}. The
younger and metal-rich stars are typically centrally concentrated; the
older and metal-poor stars are more spatially extended. In certain
cases the two populations exhibit different kinematics. This feature has been widely used for dynamical mass modelling of these galaxies
\citep{walkerPenarrubia,AmoriscoEvans,genina,hayashiCarina}.

A number of possible scenarios for the formation of two metallicity
populations have been proposed. The simplest suggest gas
reaccretion or recycling, whereby the feedback from the first episode
of star formation empties the gas reservoir and the second population
is not formed until the gas returns and cools \citep{Dong,tolstoySculptor}. These highly idealised models do not explain why the two metallicity populations should exhibit spatial segregation.

The work of \citet{elbadry} suggests baryon inflows and outflows associated with
bursts of star formation as a driving force behind stellar radial
migration. The gravitational potential fluctuations that are also responsible for
the formation of inner dark matter cores in these simulations tend to
heat the orbits of the stars over long timescales. Thus, the older and
more metal-poor stars migrate to systematically larger distances than
younger ones, creating a metallicity and an age gradient. This
mechanism would explain the presence of gradients in some dwarf
galaxies. However, unless this mechanism is coupled to another process
triggering two well separated episodes of star formation activity, it is unclear how it can lead to large spatial segregations such as that observed in, for example, Sculptor, where the effective radius of
the metal-rich population is $\sim$~0.55 that of the metal-poor
population \citep{battagliaSculptor,walkerPenarrubia}.

The works of \citet{alejandro-reion,alejandro-mergers} 
suggest gas-rich mergers as a mechanism for forming two stellar
populations that are distinct in age and spatial extent. Using the
CLUES simulations \citep{clues}, these authors find that some haloes,
ranging in stellar mass from 9$\times$10$^{6}$ to
6$\times$10$^7$~M$_{\odot}$, are able to form stars before reionization
but, due to star formation feedback and a low virial mass, they are
not able to reaccrete gas until a late time merger takes place and
funnels gas to the centre, resulting in a second burst of star
formation. A merger increases the velocity dispersion of the old stars. This scenario can explain both why the stellar
populations are distinct in age and also why the younger population is
more centrally concentrated than the older population.

In their zoom-in simulations of isolated dwarfs using the
chemo-dynamical $N$-body code \textsc{gear} \citep{gear},
\citet{jablonka} find no such late accretion events that result in the
formation of two metallicity populations. However, they do find that
the dwarf galaxies with significantly steeper metallicity gradients than others
appear in systems which have assembled \textit{early on} from
metal-poor galaxy progenitors, resulting in a very extended
distribution of metal-poor stars. $N$-body simulations show that
dwarf-dwarf mergers are not uncommon in a $\Lambda$-Cold Dark Matter
($\Lambda$CDM) universe \citep{alismergers}. Furthermore, these would
help explain photometric and kinematic anomalies present in dwarfs such as Andromeda II and Fornax
\citep{amoriscoandromeda,Lokasrotation,delpinofornaxrotation}.

Another plausible formation path has been suggested by
\citet{wright2018}. These authors identified ram pressure as a
mechanism for the reignition of star formation in field dwarf galaxies
of stellar mass in the range
9.2$\times$10$^{8}$~-~8.4$\times$10$^9$~M$_{\odot}$, and in which the star
formation history has a prolonged gap. They find that star formation
may be reignited by ram pressure due to, for example, gas blown out by
intense star formation activity from a nearby galaxy. Given
sufficiently low velocity relative to the surrounding medium, the hot gas in the outer regions
of the halo may be stripped while the gas in the inner regions may be
compressed and cooled resulting in a new star formation episode and
the formation of metal-rich stars.

\citet{kawata}, on the other hand, argue that the two metallicity populations observed in Sculptor may, in fact, be a single stellar population. These authors were able to reproduce Sculptor's steep metallicity gradient in a simulated galaxy that undergoes a single star formation episode at $z=$~13~-~20. The star formation in their simulated dwarf is powered by smooth accretion, where the gas is primarily enriched in the denser central regions, while supernovae feedback prevents star formation in the outermost regions. This mechanism creates a metallicity gradient that is sufficiently steep, such that the system appears to contain two chemodynamically and spatially distinct populations.

A successful scenario for the formation of dwarfs with two stellar
metallicity populations would need to explain both the difference in
the age/metallicity of the two populations as well as the difference
in their spatial extent. Whilst the works of \citet{alejandro-mergers}
and \citet{wright2018} have been able to achieve this, neither have
considered satellite dwarfs in a Local
Group-like environment. In this work we analyse the assembly histories of both field
and satellite dwarf galaxies with stellar masses spanning the range
10$^6$~-~10$^9$~M$_{\odot}$ in five high-resolution
hydrodynamical simulations of environments resembling that of the
Local Group. We will identify two-metallicity population dwarf galaxies and examine their
formation paths. We will further discuss how these
histories are imprinted in the observable properties of the dwarfs. 

The details of the simulations and the galaxy sample are discussed in
Section~\ref{methods}. In Section~\ref{sectionfield} we examine the
formation histories of field dwarfs with two metallicity populations and analyse the satellites in Section~\ref{sectionsat}. Section~\ref{massdepsec} explores the stellar mass dependence of these mechanisms. In Section~\ref{signaturessection} we look at the properties of the
individual stellar populations. Section~\ref{observations} examines the conditions necessary to observe bimodality in metallicity distributions and in
Section~\ref{conclusions} we summarise our conclusions. Readers who are particularly interested in dwarf satellites (for which we have uncovered a new physical process) may wish to skip directly to Section~\ref{sectionsat}.

\section{Simulations}
\label{methods}

\begin{figure*}
		\includegraphics[width=2.\columnwidth]{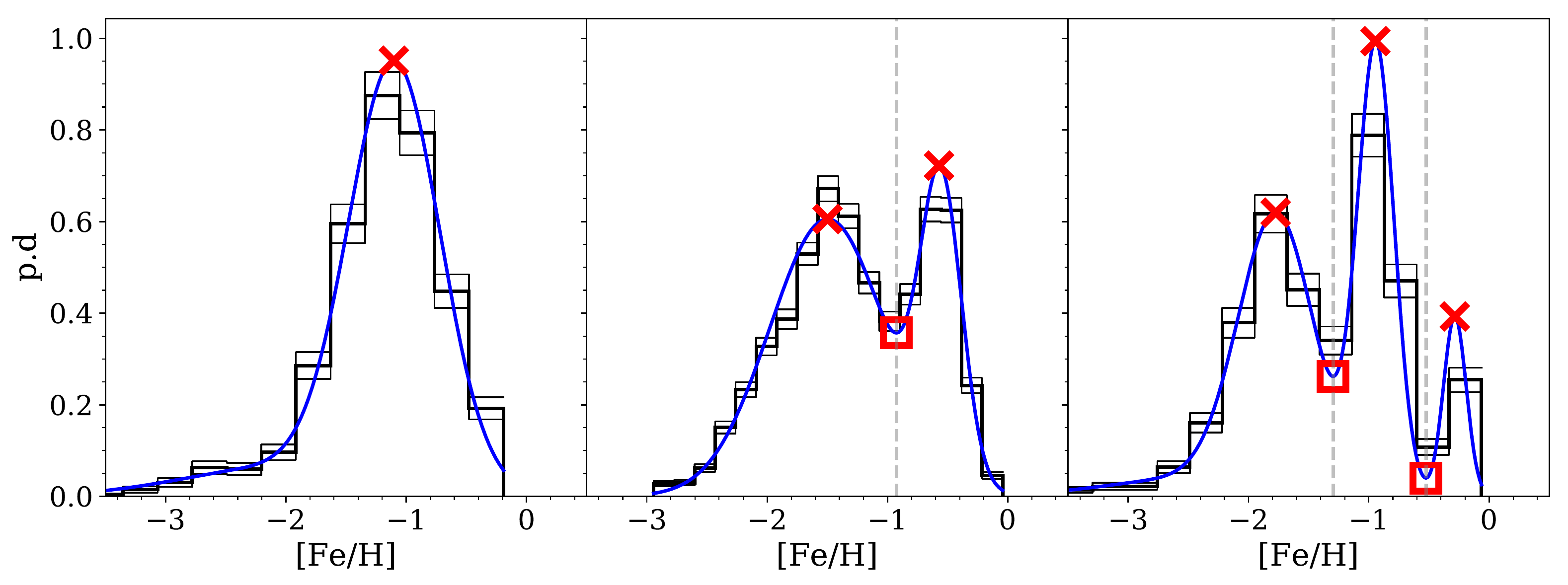}

		\caption{Metallicity probability density (p.d) for examples of galaxies with one (left), two
                  (middle) and three (right) distinct stellar
                  populations as determined by the method described in Section~\ref{defining}. The black lines show binned data and
                  corresponding Poisson errors. The blue line shows
                  the best fitting Gaussian Mixture Model. The red
                  crosses show the location of peaks; their number is
                  equal to the number of identified populations. The
                  red squares show the local minima and mark the
                  locations where populations are split (also shown
                  with a vertical dashed grey line).}

		\label{splitPlot}
\end{figure*}

\subsection{APOSTLE simulations}

A Project Of Simulating The Local Environment (APOSTLE) consists of a suite of hydrodynamical zoom
simulations of 12 cosmological volumes; volumes AP-(1-12) were
simulated at medium and low levels of resolution (L2 and L3) while
five volumes (AP-1, AP-4, AP-6, AP-10 and AP-11) were simulated at high
resolution (L1). In this work we use only the five L1 volumes. The
Milky Way and Andromeda analogues were selected from the DOVE simulation \citep{dove}. The
separations, masses, radial and tangential velocities of the halo
pairs, as well as the broad kinematics of other members of the Local
Group analogues were chosen to satisfy observational constraints
\citep{apostleAzi}. The simulations are described in detail in that
paper and in \citet{apostleTill}.

APOSTLE was run with the \textsc{eagle} code
\citep{eagleschaye,eaglecrain}, which is a modified verison of the
TreePM smoothed particle hydrodynamics (SPH) code \textsc{gadget 3}
\citep{gadget}. \textsc{eagle} was calibrated to reproduce the stellar mass function at $z$~=~0.1 in the resolved range of 10$^8$~--~10$^{11}$ M$_{\odot}$, as well as galaxy sizes and
colours. 

Radiative cooling and photoheating prescriptions in \textsc{eagle} follow
\citet{cooling}.  Star formation is stochastic and follows a pressure
law described by \citet{eagleSF}. The star formation threshold is
dependent on both the number density of hydrogen and the gas
metallicity, following the prescription of \citet{threshold}. Each
stellar particle represents a stellar population with masses
between 0.1~-~100~M$_{\odot}$, following a \citet{chabrier} initial
mass function. Newly formed stellar particles inherit their parent gas
kernel smoothed abundances. Altogether eleven chemical elements are tracked: H, He, C, N, O, Ne, Mg, Si, S, Ca and Fe. Nine of these
are tracked individually, while Ca and S are assumed to have fixed
mass ratios relative to Si. Stellar evolution and mass loss follows
the model of \citet{wiersma} and includes winds from AGB, massive
stars and SNe Type~II. The rate of SNe Type~Ia follows a time delayed
exponential distribution function. Their yields are taken from
\citet{snyields}. Lost stellar mass is distributed through an SPH
kernel to 48 gas particle neighbours. Stochastic thermal feedback from
star formation is implemented following \citet{feedbackdallaveccia}. Hydrogen
reionization is modelled by a spatially-uniform and time-dependent
ionizing background, which is turned on instantaneously at $z$~=~11.5
\citep{reionization,eagleschaye}.

\subsection{Constructing merger trees}

We constructed merger trees using the \textsc{HBT+} halo finder
\citep{hbt}. Each halo in the simulation is assigned a `track ID'
which is used to follow the main progenitor across time. Thus, the
position and velocity of the main progenitor at every snapshot in the
simulation may be determined. We calculate the centres of subhaloes
using the ``shrinking spheres" algorithm \citep{power} on at
least 100 stellar particles, if present, or on the bound dark
matter particles if not.

\subsection{Calculating V-band luminosities}

The V-band luminosities for our simulated dwarfs are calculated following the \textsc{galaxev} population synthesis model of \citet{stellarsynthesis}. The model provides absolute $\mathrm{AB}$ magnitudes of a stellar population with mass M$_{\odot}$, age $t_*$ and metallicity $Z_*$, assuming a \citet{chabrier} initial mass function. The magnitudes are tabulated on a grid of metallicity (1$\times10^{-4}$~<~$Z_{*}$~<~5$\times10^{-2}$) and age (0.1~<~$t_{*}$/Myr~<~2$\times$10$^4$). For each simulated stellar particle, described by its initial mass, age and metallicity, we thus perform a 2-dimensional interpolation over the age-metallicity grid and obtain its absolute magnitude in the V-band. If the simulated stellar particle falls outside the metallicity range of interpolation, we assume that its metallicity is either the minimum or maximum value of the tables (i.e. we do not perform an extrapolation).

In this work we will use these luminosities to calculate quantities that are compared to observations (such as the half-light radius).

\subsection{Defining stellar metallicity populations in galaxies}
\label{defining}

We now define what we mean by galaxies with two metallicity
populations. 

We use [Fe/H] as the measure of metallicity, as is common
in observational work\footnote{The adopted mass fractions are 0.0014M$_{\odot}$ for iron and 0.7381M$_{\odot}$ for hydrogen.}. For each stellar particle, kernel-smoothed metallicity abundances are used. In order to identify the number of stellar metallicity populations in
dwarf galaxies we employ a peak finding algorithm. We fit the [Fe/H] distribution with
a Gaussian Mixture Model (GMM) \citep{gmm}  in order to eliminate the effects of noise in the distribution. We exclude extremely
metal-poor stellar particles ([Fe/H] < -4) from the fit as these lie in a long tail of the distribution that may bias the fit. We have
verified that this choice indeed removes the extremely low-metallicity
tail and does not discriminate against identifying the more metal-poor
subpopulations\footnote{Extremely metal-poor stars typically make up 0.5 per cent of [Fe/H] distributions and we assign these stars to the metal-poor population in further analysis.}. 

To identify the best performing GMM, we evaluate $\chi^2$ for the
binned data, with the optimal number of bins determined by Doane's
formula for non-normal distributions \citep{doane}. We test $n$
components for the GMM, where $n < m/3$, $m$ being the number of bins,
and evaluate their quality in representing the data using the Akaike
Information Criterion corrected for sample size
\citep[AICc;][]{akaike}. The model with the smallest AICc is selected
and a peak finding algorithm is run on the resultant metallicity
distribution function. A peak location is defined as a point, $i$,
where the function, $f$, has a local maximum, such that
$f_{i - 1} < f_{i} > f_{i + 1}$. The number of identified peaks in the
distribution of [Fe/H] is the number of unique stellar populations in
the galaxy. Once the peaks have been found, in cases where the number
of populations is greater than one, we split the stellar particles
into their respective populations by applying a hard cut at the
positions of the local minima between the peaks.

By definition, our method identifies galaxies with two or more peaks in their
stellar metallicity distributions. However, in some cases the size of
a subdominant population may be negligible, such that the galaxy
effectively consists of a single stellar population. We thus consider
dwarfs in which the mass ratio of the smaller to the larger population
is less than 10 per cent as single population galaxies. We additionally limit our sample to galaxies where any stellar population contains at least 100 stellar particles. At this limit sampling noise becomes important, although some galaxy properties can be measured with 20 per cent accuracy \cite[see the Appendix of][]{genina} and, provided that the metallicity distribution exhibits distinguishable peaks, it is still possible to separate different stellar populations.

In Fig.~\ref{splitPlot}, we show [Fe/H] distributions for examples of galaxies with one, two and three
identified stellar populations. The \textsc{eagle} model produces a
wide range of stellar metallicity distributions and it is clear that the method works well
in identifying populations of varying sizes in the distributions of [Fe/H]. We note that the metallicity distributions may be sensitive to the specifics of subgrid physics. In particular, metal mixing is not properly accounted for in our simulations, although the use of SPH-smoothed metallicities does somewhat mitigate this issue \citep{apostleMetal}. Nevertheless, the origin of metallicity distribution bimodality (see Sections A1 and A2 of the Appendix), spatial segregation and, thus, the main results of this work should not be affected. 

The resulting sample consists of 290
galaxies with $\sim$~43~per~cent containing two metallicity
populations (61 field and 64 satellite dwarfs), 10~per~cent containing
three (16 field and 14 satellites) and 47~per~cent containing a single
stellar component (66 field and 69 satellites). According to these statistics, nearly half of all simulated dwarfs above $\sim$10$^6$~L$_{\odot}$ exhibit metallicity bimodality in their [Fe/H] distributions. We find these ratios to be approximately constant across the range of luminosities above $\sim$4$\times$10$^6$~L$_{\odot}$. Below this value the fraction of two-population dwarfs decreases steadily down to 20 per cent, but this is due to the imposed minimum number of stellar particles for a given metallicity population. We will discuss observational implications of this in Section~\ref{observations}.

In what follows we consider only the dwarfs that contain two stellar
populations, with a particular emphasis on the formation mechanisms responsible for the \textit{spatial segregation} within these galaxies. We leave the three-population galaxies for future work. These may be interesting in the context of the dynamical method of
\citet{walkerPenarrubia} and may allow additional constraints to be
placed on the slope of the dark matter density profiles.

\subsubsection{Properties of our galaxy sample}

\begin{figure}
       
		\includegraphics[width=1\columnwidth]{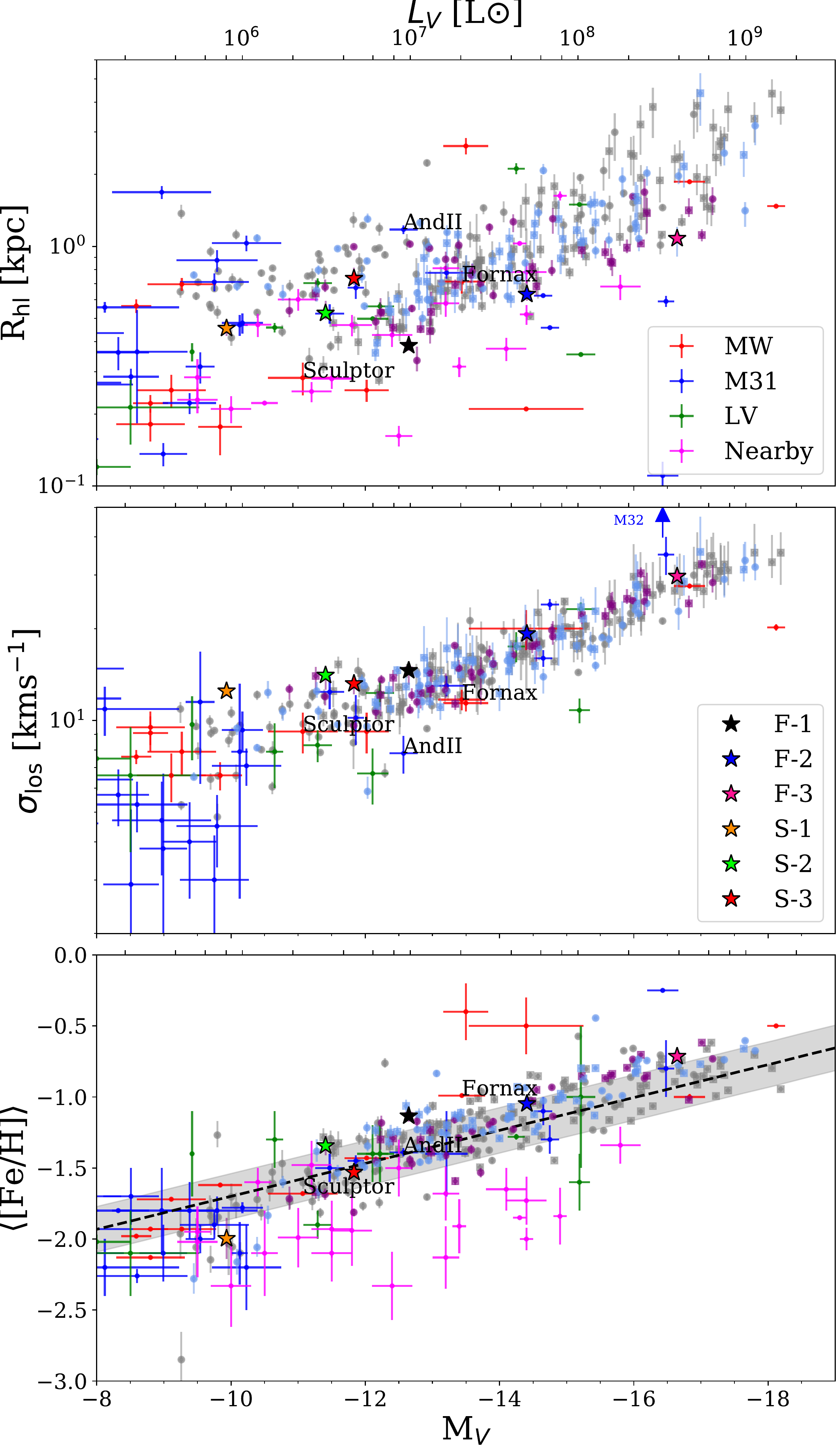}
		
		\caption{\textit{Top:} Projected half-light radii for our sample of dwarfs as a function of M$_V$, with error bars representing the 16$^{\mathrm{th}}$ and the 84$^{\mathrm{th}}$ percentiles.  Model galaxies with
                  two metallicity populations are shown with blue
                  symbols, with satellites and field dwarfs
                  represented by circles and squares,
                  respectively. The remainder of the sample is shown with grey symbols. The data for the Local Volume and nearby galaxies from \citet{mcconnachieCensus} is shown with symbols
                 identified in the legend.  Corresponding $V$-band luminosities are shown in the upper horizontal axis.  \textit{Middle:} The $V$~-~band luminosity-weighted line-of-sight velocity dispersion for galaxies in our sample as a function of M$_{V}$.  \textit{Bottom:} Mean luminosity-weighted stellar metallicity as a function of
                  M$_V$ for our sample of dwarfs. The black dashed line shows the relation derived by \citet{kirbyrelation}. The specific examples of field and satellite galaxies that will be discussed later on are marked across the three panels with stars, as identified in the middle panel. }

\label{magnitude}
\end{figure}

We first demonstrate that the dwarfs in our sample have basic structural properties that resemble those of real field and satellite dwarfs. In Fig.~\ref{magnitude} we show the projected half-light radii, $\mathrm{R_{hl}}$, line-of-sight velocity dispersion, $\mathrm{\sigma_{los}}$, and the mean mass-weighted stellar metallicities, $\mathrm{\langle[Fe/H]\rangle}$, as a function of $V$-band magnitude, M$_{\mathrm{V}}$. We compare the values for our sample of galaxies with observed Local Group
dwarfs. The Local Group data have been taken from \citet{mcconnachieCensus} and includes all dwarfs for which measurements of half-light radius, line-of-sight velocity dispersion and metallicity are available. Note that some of these measurements do not have error bars. The Milky Way and Andromeda satellites (red and blue symbols, respectively) as well as other Local Volume dwarfs (green symbols) shown in Fig.~\ref{magnitude} are the same across the three panels. The velocity dispersion and half-light radii for the simulated dwarfs are shown with error bars representing the 16$^{\mathrm{th}}$ and the 84$^{\mathrm{th}}$ percentiles found by generating 1536 isotropically distributed lines of
sight \citep{healpix}. The error bars for mass-weighted mean stellar metallicities are calculated by taking 1000 samples of stellar particles with the number of particles within each sample, $N_{\mathrm{*,sample}}\sim N_{\mathrm{*,total}}/10$. We exclude stellar particles with [Fe/H]~>~-4 from the mean metallicity calculation, as these metallicities are sensitive to the mass resolution of our simulations \citep{eagleschaye, apostleMetal}. The two-population galaxies are shown in blue, satellites are represented with circles and field dwarfs with squares. It can be seen that the two-population galaxies occur within a wide range of luminosities and galaxy sizes.

As a result of the imposed minimum number of stellar
particles, our simulated galaxies are only comparable to some of the
brightest, or classical, Local Group dwarfs. The black dashed line in the bottom panel of Fig.~\ref{magnitude} displays the mean-metallicity~--~luminosity relation derived by \citet{kirbyrelation}, extrapolated to higher luminosities. The grey bands represent the scatter. Our simulated dwarfs trace this relation. 

Sculptor, Fornax and Andromeda~II, particularly well studied examples of two-population dwarfs, are identified on the plot. It can be seen that our sample contains galaxies of comparable luminosities to these three dwarfs. A number of simulated galaxies match the size and velocity dispersion measurements for Fornax, although its mean metallicity is above the locus traced by our galaxy sample. The sample contains matches for Sculptor in velocity dispersion and mean metallicity, although the half-light radius of Sculptor is below those seen in our simulations at comparable luminosities. The size-luminosity relation in our simulations flattens at low luminosities ($\lesssim$10$^6$L$_{\odot}$). As discussed in more detail in \citet{campbell}, the galaxy sizes are sensitive to spatial resolution and are typically larger than 2.8 times the Plummer-equivalent softening of our simulations (i.e. $\sim$~0.4~kpc), which is the radius above which the forces are Newtonian. We have also explicitly checked that the luminosities derived from the stellar synthesis models are not responsible for the apparently large sizes of these galaxies, as their half-mass and half-number radii have comparable values. The sizes of these low-luminosity dwarfs and those of their subpopulations should be taken with caution in the rest of this work, but should not affect our ability to identify spatial segregation and its origin.

We now identify a subsample of these simulated dwarfs that also exhibit significant spatial segregation, comparable to that derived for Sculptor and Fornax, and examine their formation mechanisms.

\section{Assembly history of field dwarfs}

\label{sectionfield}

\begin{figure}
	
		\includegraphics[width=\columnwidth]{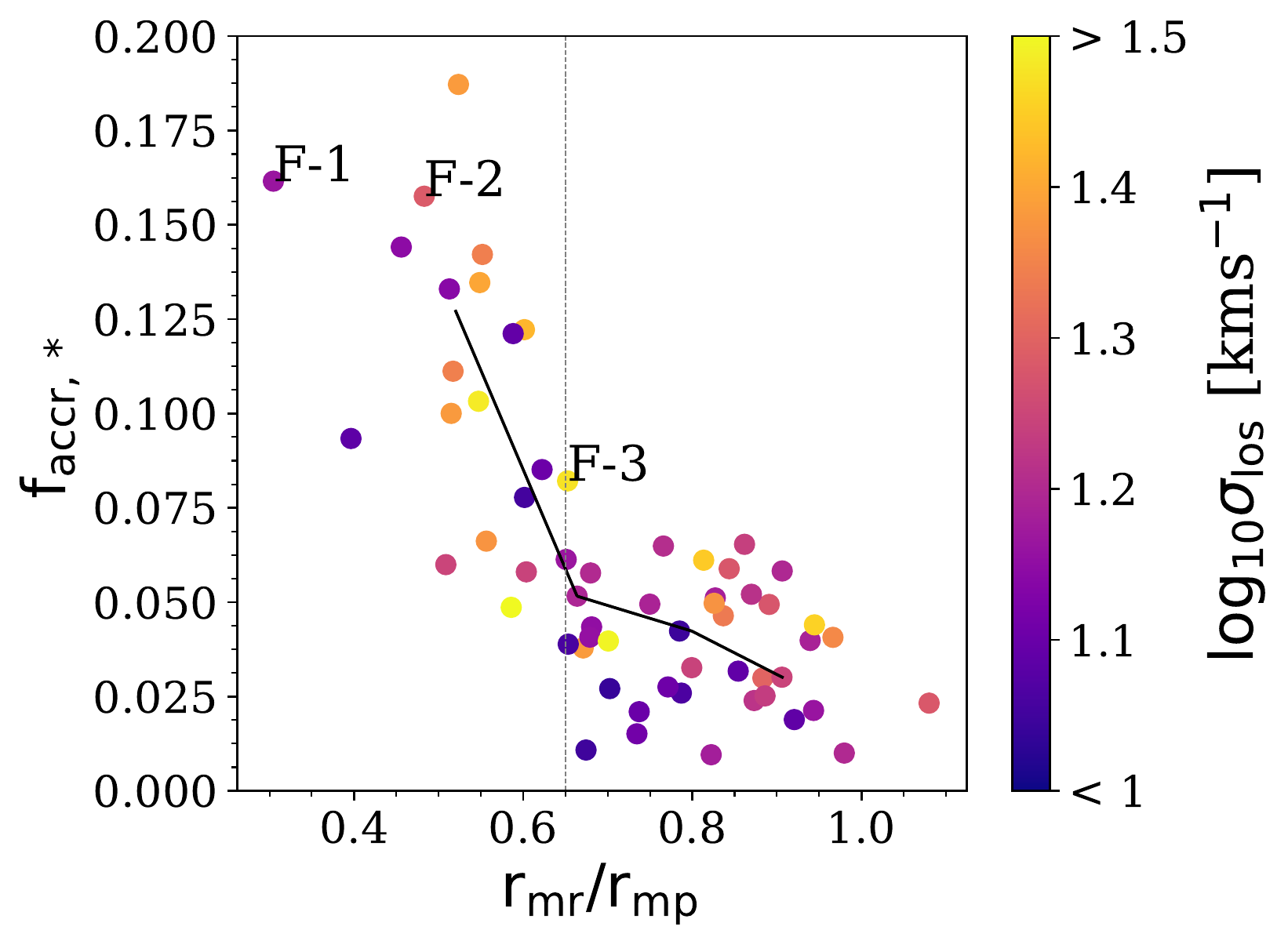}
                \caption{The fraction of the accreted
                  stellar particles, $\mathrm{f_{accr,*}}$, as a
                  function of spatial segregation for galaxies with two metallicity populations. The points are coloured by the median of the line-of-sight velocity dispersion. The grey
                  vertical dashed line separates dwarfs with low and
                  high $\mathrm{r_{mr}/r_{mp}}$. The galaxies on the left of this line are selected as our subsample of well spatially segregated galaxies. The black solid line
                  shows the median relation in four bins of equal
                  galaxy numbers. Three galaxies that will be studied in particular detail are labelled F-(1-3).}
\label{accretionFig}
\end{figure}

It has previously been suggested by \citet{alejandro-mergers} and
\citet{jablonka} that mergers play a significant role in the formation of
two-population systems. A merger provides a natural explanation
for both the large spatial extent of the metal-poor stellar population
and the reignition of star formation activity that results in the
formation of a more metal-rich population. If this is indeed the case,
simulated systems with strong spatial segregation between two populations should also exhibit larger
fractions of accreted material. Fig.~\ref{accretionFig} shows the fraction of stellar
particles that have formed outside $\sim$0.15$~\times~$R$_{200}$ of the
main progenitor\footnote{We define R$_{200}$ as the radius enclosing a
  mean density of 200 times the critical density of the Universe. We
  find the 0.15$\times$R$_{200}$ cut to be a reliable tracer of the
  accreted stellar particles since the merging galaxies tend to form a
  large fraction of stars during the merging process itself,
  i.e. within R$_{200}$. In practice, we vary this factor between
  (0.1-0.4)$\times$R$_{200}$ to better suit the stellar halo size of
  each system.} (i.e that have been accreted), $\mathrm{f_{accr,*}}$,
as a function of $\mathrm{r_{mr}/r_{mp}}$, the ratio of the metal-rich to the metal-poor half-mass radii, with the colours representing the median of the line-of-sight velocity dispersion in these systems as a tracer of mass. It is evident that the two-population dwarfs make up two subsets: those that accrete a significant fraction of their stars, typically $\mathrm{f_{accr,*}}>0.05$, and in which the spatial segregation is large ($\mathrm{r_{mr}/r_{mp}}$~$\lesssim$~0.65) and those where both the amount of accretion and the spatial segregation are small. This separation does not appear to be dependent on the mass of these systems as large spatial segregation is seen in both high ($\sigma_{\mathrm{los}}\sim25$~kms$^{-1}$) and low velocity dispersion systems ($\sigma_{\mathrm{los}}\sim10$~kms$^{-1}$). We thus select dwarfs with $\mathrm{r_{mr}/r_{mp}}<0.65$ as our subsample of galaxies that have two spatially segregated populations, which are a main focus of this work.

\begin{figure}

		\includegraphics[width=0.9\columnwidth]{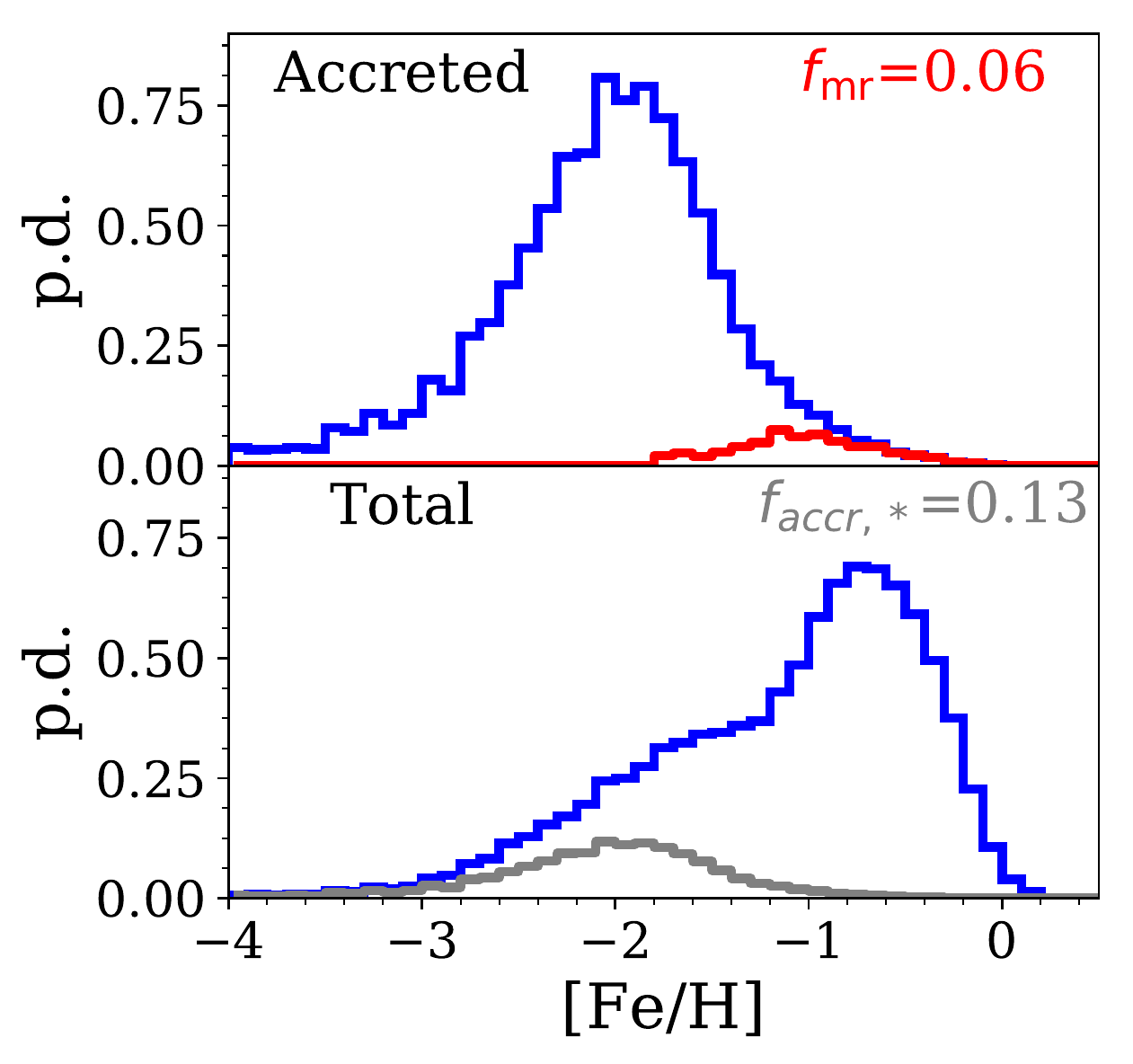}
                \caption{\textit{Top:} Histogram of metallicities of
                accreted particles in field galaxies with
                $\mathrm{r_{mr}/r_{mp}}$~<~0.65 (blue). The red
                histogram corresponds to the subset of accreted
                particles that belong to the metal-rich
                population. \textit{Bottom:} Histogram of all
                stellar particles in blue. The subset of particles that are accreted are shown in grey. The histograms give equal weight to each galaxy.}
\label{feh_accrFig}
\end{figure}

The top panel of Fig.~\ref{feh_accrFig} shows the distribution of the
metallicities of accreted particles for all field dwarfs with two spatially
segregated populations, with each galaxy given equal weight. The subset of accreted stellar particles that belong to the metal-rich population of their host galaxy is shown by the red histogram. It is evident
that only a very small fraction of the accreted stars belong to the metal-rich population
(typically $\sim$6 per cent). We thus conclude that the metal-rich
population in these dwarfs is primarily formed {\em in-situ}. The bottom
panel shows the distribution of all particles in
galaxies with two well-segregated populations (blue), with equal weight per galaxy. It appears that the field dwarfs show a preference for a more dominant metal-rich population. The grey histogram shows a subset of particles that were accreted. These particles are partly responsible for the metal-poor
bump seen in two-metallicity population galaxies, though typically only $\sim$~0.13 of the
total stellar mass is made up of the stars that were accreted. Nonetheless, a large accretion fraction suggests prior merger activity; processes associated with these mergers are capable of
inducing spatial segregation, as we now discuss.

\begin{figure*}
		\includegraphics[width = 2\columnwidth]{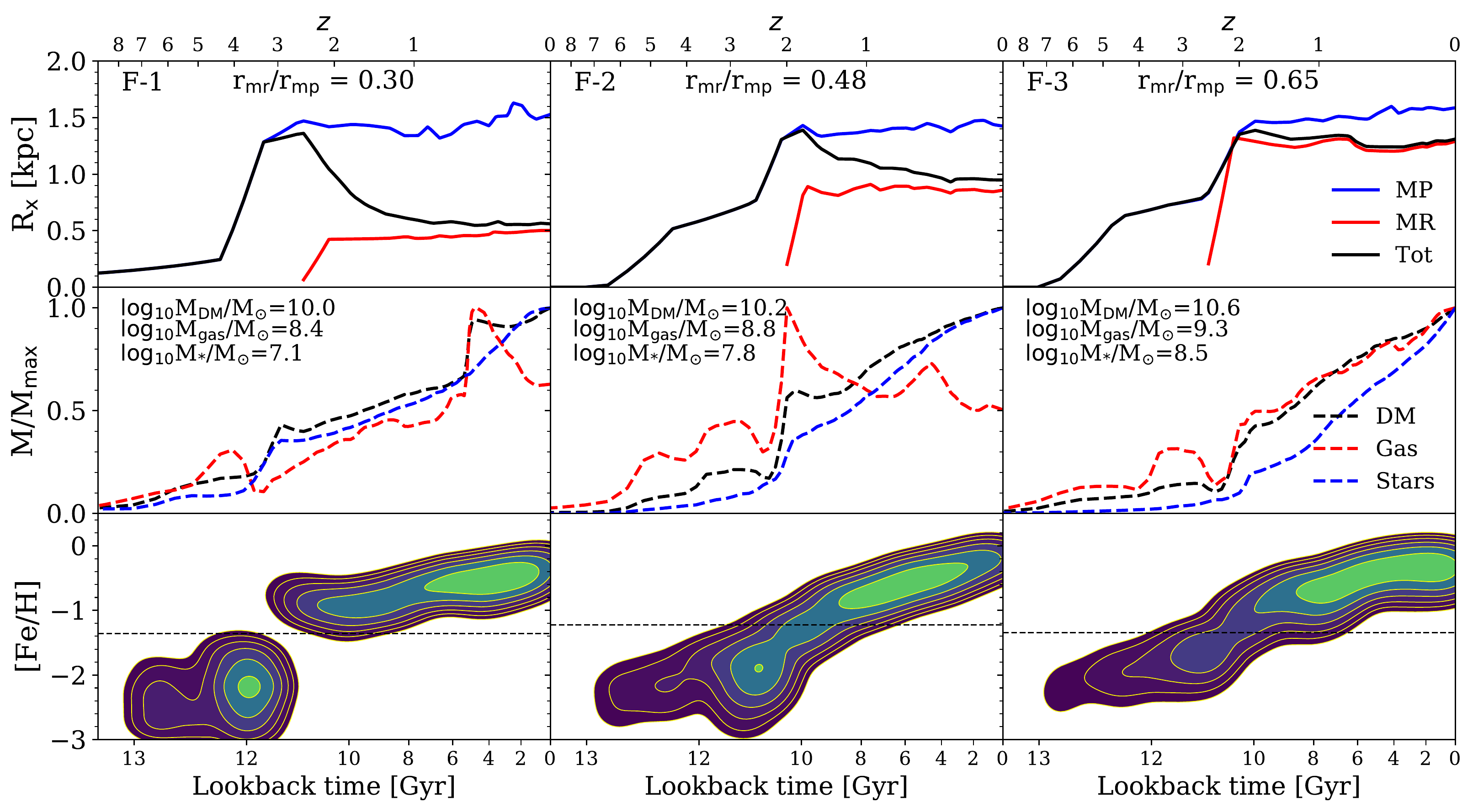}

		\caption{\textit{Top:} The evolution of the half-mass radius of the
                  metal-poor stars formed {\em in-situ} (blue), the
                  metal-rich stars (red) and the entire set of stellar particles formed \textit{in-situ} (black). The lookback time is
                  shown on the bottom axis and the redshift on
                  the top. The subpopulations are tracked from the
                  time when they first contain at least 50 stellar
                  particles. From left to right are three examples of
                  galaxies with progressively poorer spatial
                  segregation. \textit{Middle:} The evolution of mass for the dark matter (black), gas (red) and stars (blue). Sudden increases in dark matter mass correspond to mergers. The mass is normalised by the maximum component mass, shown by the labels in the upper left corners. Note that galaxy F-2 merges with two haloes at approximately the same time. \textit{Bottom:} The metallicity-age relation for the three example galaxies. The rise in metallicity is seen to occur following a merger. The contours in probability density are spaced geometrically by factors of 2 and are consistent for the three galaxies.}

		\label{figmerger}
\end{figure*}

\subsection{Spatial segregation through mergers}

We now investigate how mergers influence the spatial extent of
the two metallicity populations. In Fig.~\ref{figmerger} we illustrate this by
tracking the evolution of the half-mass radii of three individual
galaxies, ranging from the largest spatial segregation ($\mathrm{r_{mr}/r_{mp}}=0.30$) to the segregation at which we cut to select our subsample ($\mathrm{r_{mr}/r_{mp}}=0.65$). In the top row, the blue line represents the stars belonging to
the metal-poor population formed {\em in-situ} and the red line shows the metal-rich population. The black line is the \textit{total}
half-mass radius. The half-mass radii are tracked from the moment when
50 stellar particles are present within the virial radius of the dwarf. We exclude the accreted stars as these make up a small fraction of the overall stellar population and their half-mass radii, throughout a dwarf's history, are not easily defined. The middle row shows the evolution of dark matter (black), gas (red) and stars (blue) bound mass\footnote{Bound mass is determined via a core-averaged unbinding procedure, as implemented in \citet{hbtoriginal, hbt}.}, normalized by their maximum mass throughout the history of the dwarf, $\mathrm{M/M_{max}}$. The logarithmic maximum masses are shown in the upper left of each plot. The presence of mergers is evident as a sudden increase in dark matter mass. The metallicity-age distribution is shown in the bottom row for stellar particles present within the galaxy at $z=0$. The black dashed line displays the value of the metallicity at which we split the two populations. 

It can be seen that the mergers at early times are associated with a large increase in the half-mass radius of the system. The metal-poor particles that were formed \textit{in-situ} prior to the merger move to larger characteristic radii. This is likely due to redistribution of particle energies as a result of a rapidly changing gravitational potential. This effect is particularly evident in dwarf F-1, where the half-mass radius almost quadruples in size. Dwarf F-1 undergoes a second major merger at $\sim$~5~Gyr lookback time, with a secondary-to-primary halo mass ratio of $\mu\sim$~0.3. This mass ratio is far smaller than that at $\sim12$~Gyr ($\mu\sim$~1) and has little effect on the spatial segregation of the two populations. 

The early mergers in the three systems coincide with an increase in the number of stars formed and are followed by a significant increase in the stellar metallicity. As we demonstrate in Section~\ref{dip} of the Appendix, these mergers are associated with a rise in star formation activity followed by a steep drop, as large fractions of gas are expelled in winds. The enrichment of the interstellar medium from the newly formed stars continues during this period of low star formation activity and the metal-rich stars are consequently formed at systematically higher metallicities. The metal-rich population proceeds to form gradually until late times, creating a significant gap between the peak metallicities of the two populations (see Section~\ref{gap} of the Appendix). We examine the origin of metallicity distribution bimodality and where the gas particle enrichment occurs in greater detail in the Appendix. 

It can be seen in Fig.~\ref{figmerger} that the half-mass radius of the metal-poor population increases to $\sim$~1.5~kpc in all three dwarfs following the merger and the characteristic radii at which the metal-rich stars form appears to determine the extent of spatial segregation between two metallicity populations in these systems. From left to right of Fig.~\ref{figmerger} the dwarfs increase in stellar and halo mass and the size of the metal-rich population in these galaxies likely follows a mass-size relation (see the top panel of Fig.~\ref{magnitude}). For instance, like F-2, dwarf F-3 undergoes an early merger; yet the spatial segregation between its two populations is smaller. This galaxy retains a larger fraction of
gas following the merger, as well as accreting new gas, such that star formation continues during and immediately after the merger. This results in continuous enrichment of the stellar particles and a less pronounced gap in the distribution of metallicities. For a more massive halo, new stars are forming at correspondingly larger radii, limiting the spatial
segregation. Nonetheless, the merger that this galaxy
experienced increased the extent of the metal-poor population
sufficiently that the final spatial segregation between the two
populations is significant.

Nonetheless, as previously seen in Fig.~\ref{accretionFig}, the spatial segregation is not determined by the final mass of the system, represented by velocity dispersion in that figure, and the radii at which the metal-rich stars form are additionally affected by halo assembly history, gas supply and environment.

\begin{figure*}
    \centering
		\includegraphics[width=1.3\columnwidth]{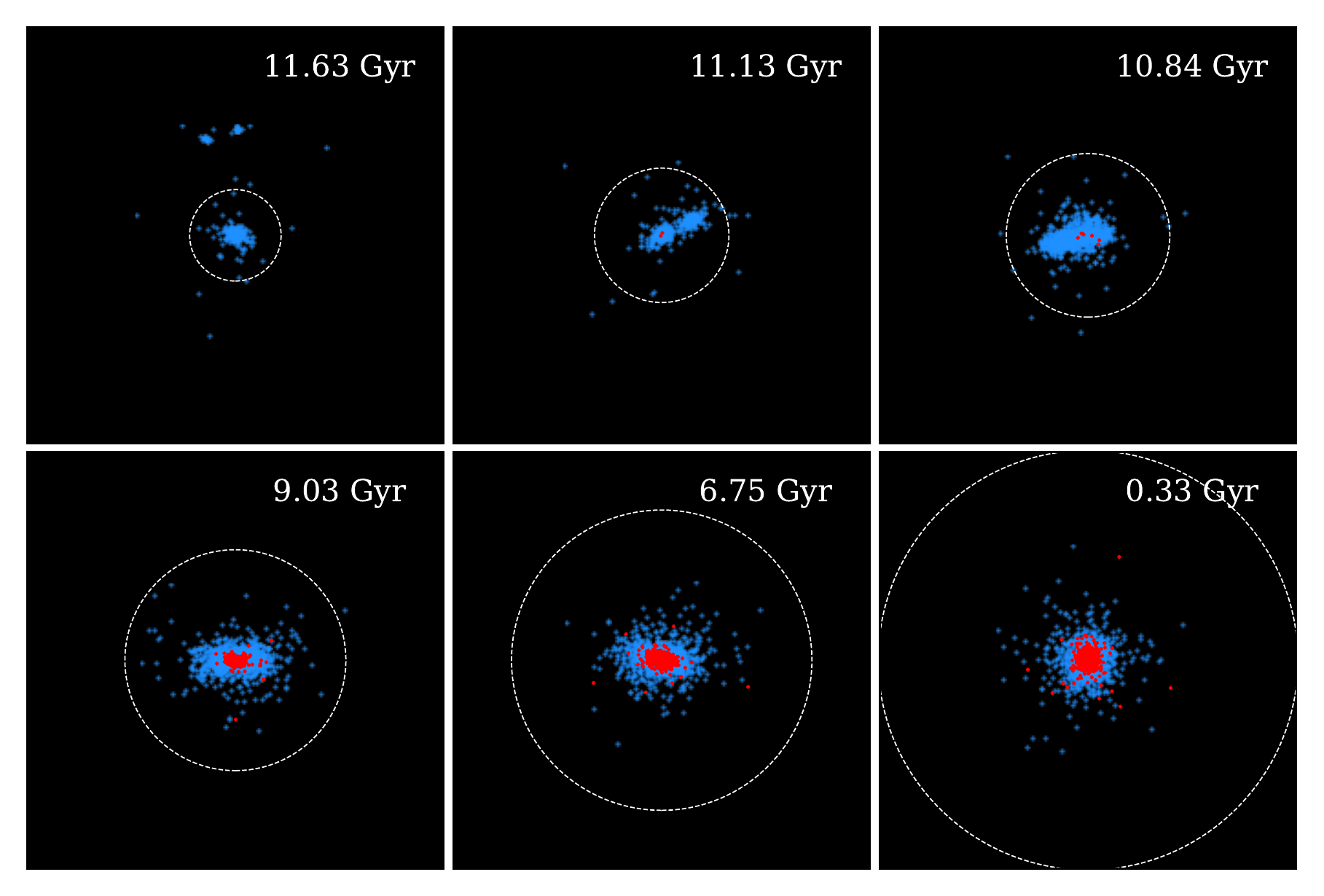}
		\includegraphics[width=1.3\columnwidth]{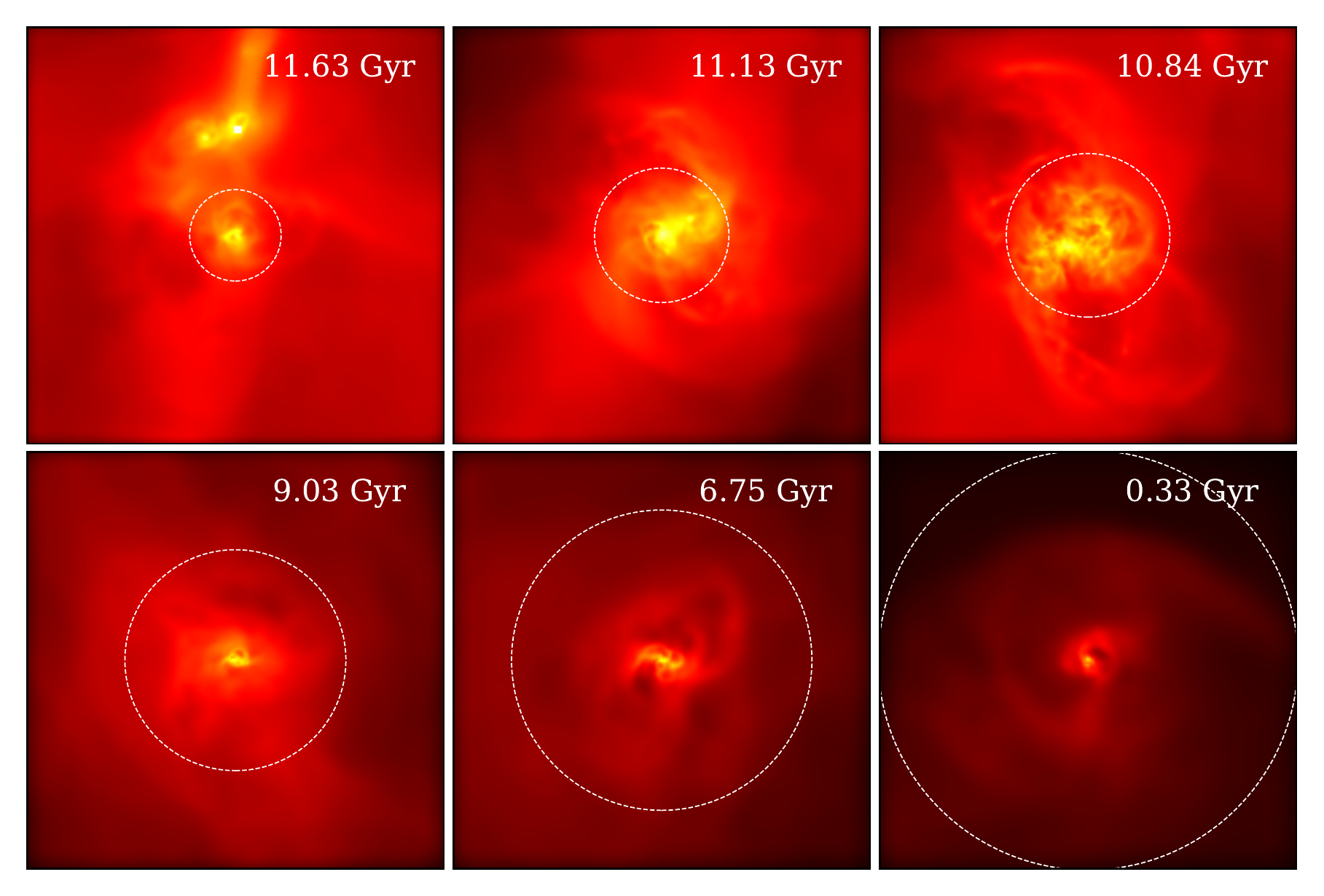}
        
\includegraphics[width=1.3\columnwidth]{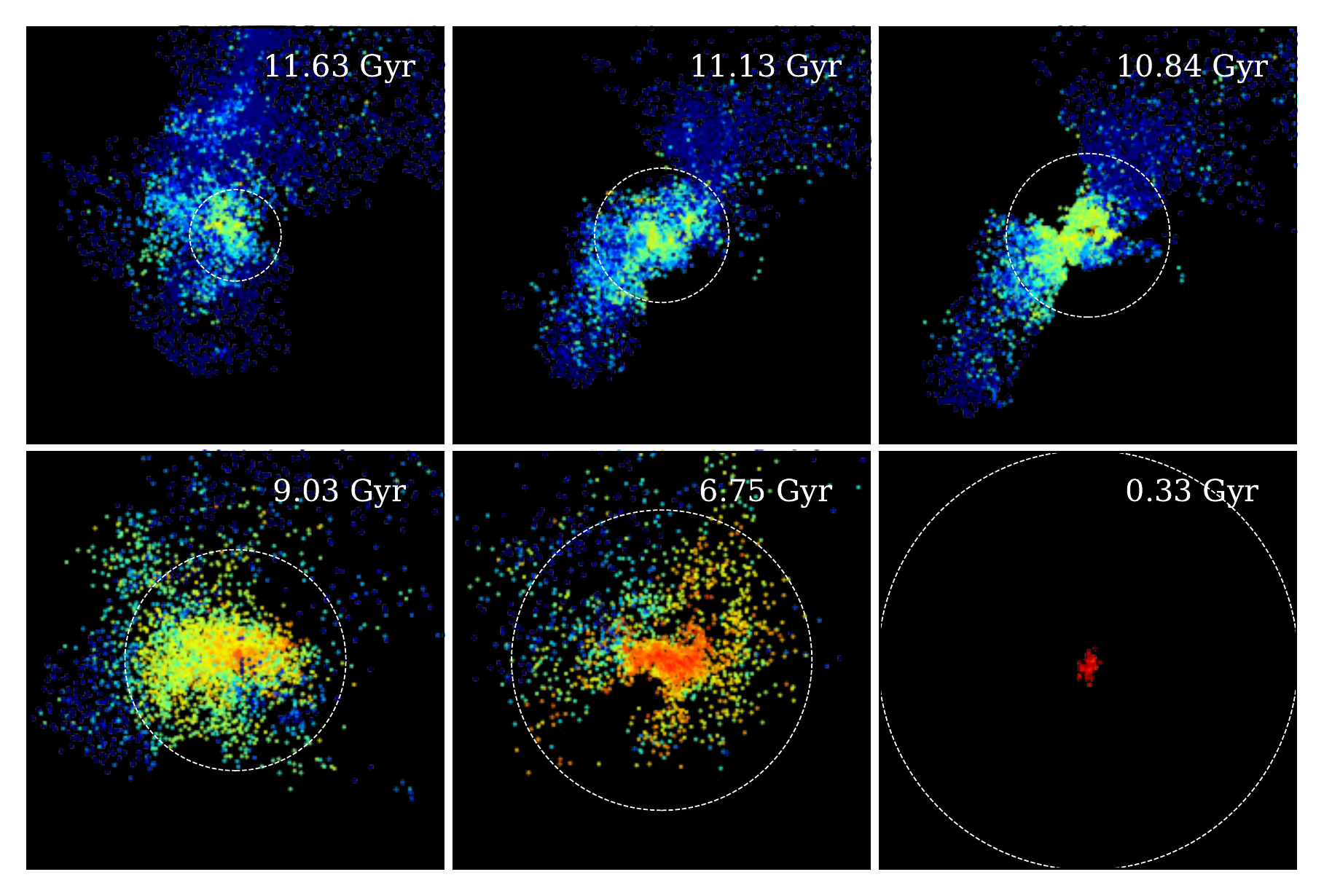}        
        
\caption{A field dwarf (labelled F-2) at 6 stages of its
  evolution. Lookback time is shown in the top right of each image. The images cover a cube of 50 kpc on a
  side. The galaxy is shown from a direction perpendicular to its
  angular momentum vector at each time. The white circles show the virial
  radius. \textit{Top:} stellar particles with metal-poor particles in
  blue and metal-rich ones in red. \textit{Middle:} the gas content of the
  dwarf. The kernel smoothed gas density varies from low density (dark
  red) to high density (yellow). \textit{Bottom:} gas particles
  destined to form stars, coloured by their metallicity from lowest
  (blue) to highest (red). }

		\label{fieldmovie}
\end{figure*}

\subsection{Merger-induced formation of two metallicity populations}
\label{mergerexamples}

We have seen that the spatial segregation between two
metallicity populations in field dwarfs is related to mergers and their effect on the size of the metal-poor population as well as on the radii at which the metal-rich stars form. We now investigate the role that mergers play in the formation of two metallicity populations further.

In Fig.~\ref{fieldmovie} we show the evolution of
one particular two-population field dwarf with a strong spatial
segregation ($\mathrm{r_{mr}/r_{mp}}$~=~0.48), previously labelled as
F-2. The top panel shows the stellar
particles in the galaxy; the metal-poor stars are coloured blue and
the metal-rich stars red. The middle panel displays the corresponding
gas densities on a logarithmic scale, with black indicating the lowest
densities and yellow the highest. The bottom panel shows the gas
particles destined to form a star within the galaxy, coloured by their metallicity at that time, with blue the most metal-poor and red the most
metal-rich.

At the lookback time of 11.63~Gyr the main progenitor is approached by two gas-rich
haloes. These haloes will supply some of the gas that will form stars
within the galaxy. By 11.13~Gyr those haloes have combined into a
single halo that merges with the main progenitor at around
10.84~Gyr. The mass ratio in dark matter is $\mu\sim1$, making this a major merger. The virial radius of the system, R$_{200}$
(indicated by a white dashed circle), has visibly increased following
the merger as has the extent of the metal-poor stellar
population. During the merger, at 10.84~Gyr, one can see gas bubbles
extending beyond the virial radius, the result of a wind powered by
the large amounts of energy released by supernovae from stars that
formed as a result of the merger. This gas is expelled beyond the
virial radius and the majority will never return to form stars. The first
metal-rich stars have been formed in this burst of star formation. A
small fraction of gas remains at 9.03~Gyr and down to very late times
this gas reservoir is enriched within the galaxy and continuously
depleted by further formation of metal-rich stars.

Dwarfs F-1 and F-3 follow a similar formation history with the key differences being the halo masses and the ability to retain gas immediately after the merger. It is thus evident that two-population systems that end
up with small values of $\mathrm{r_{mr}/r_{mp}}$ follow a well defined
formation mechanism. Dwarfs with $\mathrm{r_{mr}/r_{mp}}$~<~0.65 can
form within a wide range of halo and stellar mass. In these galaxies the
peak of star formation occurs simultaneously with the merger,
typically between 9-12 Gyr in lookback time. The associated burst of
star formation leads to the expulsion of large amounts of gas, with
only the most strongly bound gas remaining. The gas expulsion causes a
decrease in star formation activity and, therefore, a drop in the
number of metal-poor stars formed. The stars formed during the merger
proceed to enrich the interstellar medium from which the metal-rich
stars form, giving rise to a gap in the metallicity distribution. The
dwarfs that do not follow these conditions typically end up with
weaker spatial segregation and less clear bimodality in the
metallicity distribution.

\section{The formation of two populations in satellite dwarfs}
\label{sectionsat}

\begin{figure}
	
		\includegraphics[width=\columnwidth]{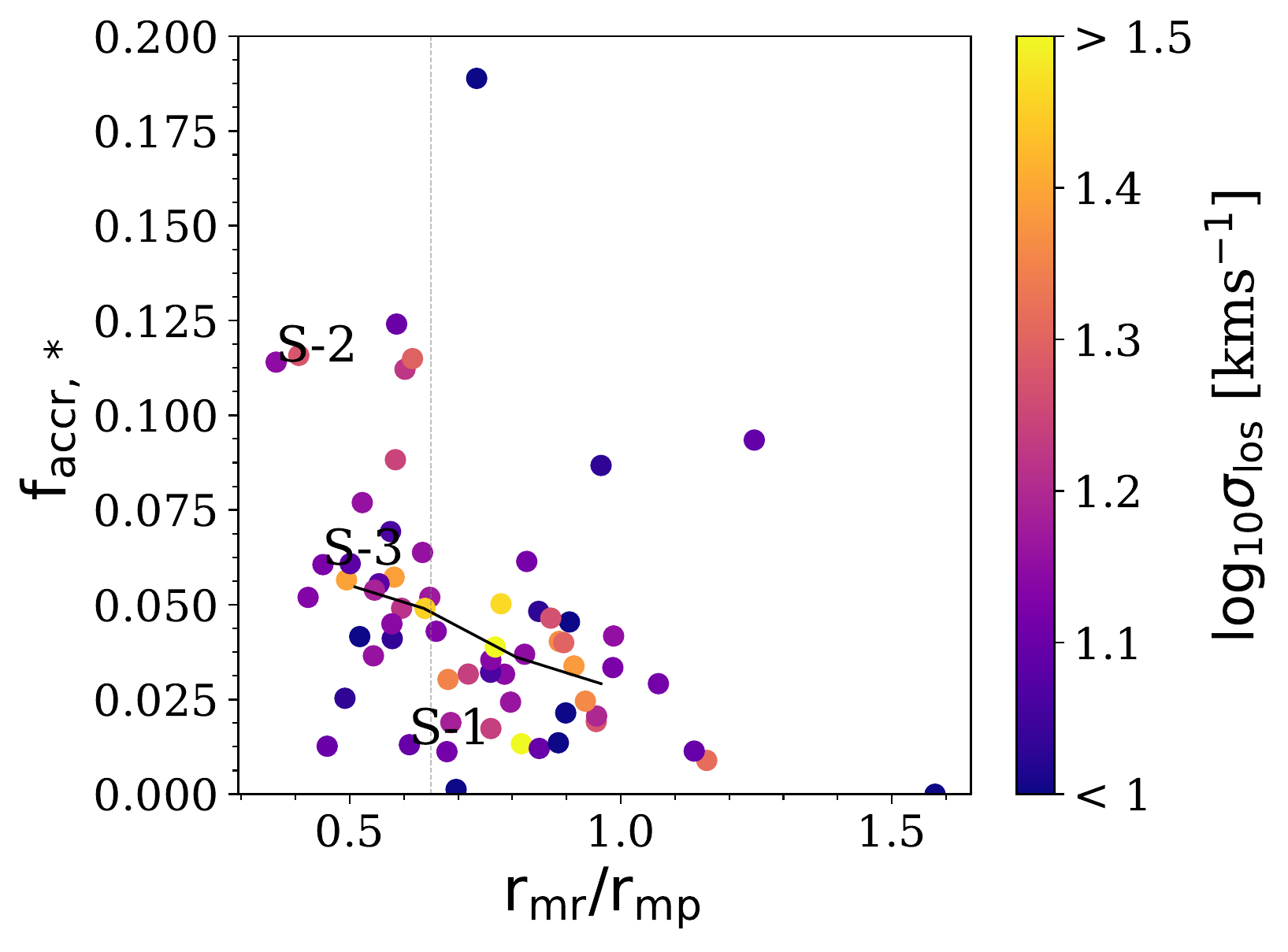}

		\caption{Similar to Fig.~\ref{accretionFig}, but for
                  satellite dwarfs: the stellar
                  accretion fraction for satellite dwarfs. The
                  two-population satellites are shown with blue
                  circles. The vertical grey line separates well- and
                  poorly spatially segregated metallicity
                  populations. The black line shows the median of the
                  relation. Three dwarfs that will be later discussed in particular detail are labelled S-(1-3).}
		\label{sataccr}
\end{figure}

\begin{figure}
	
		\includegraphics[width=0.9\columnwidth]{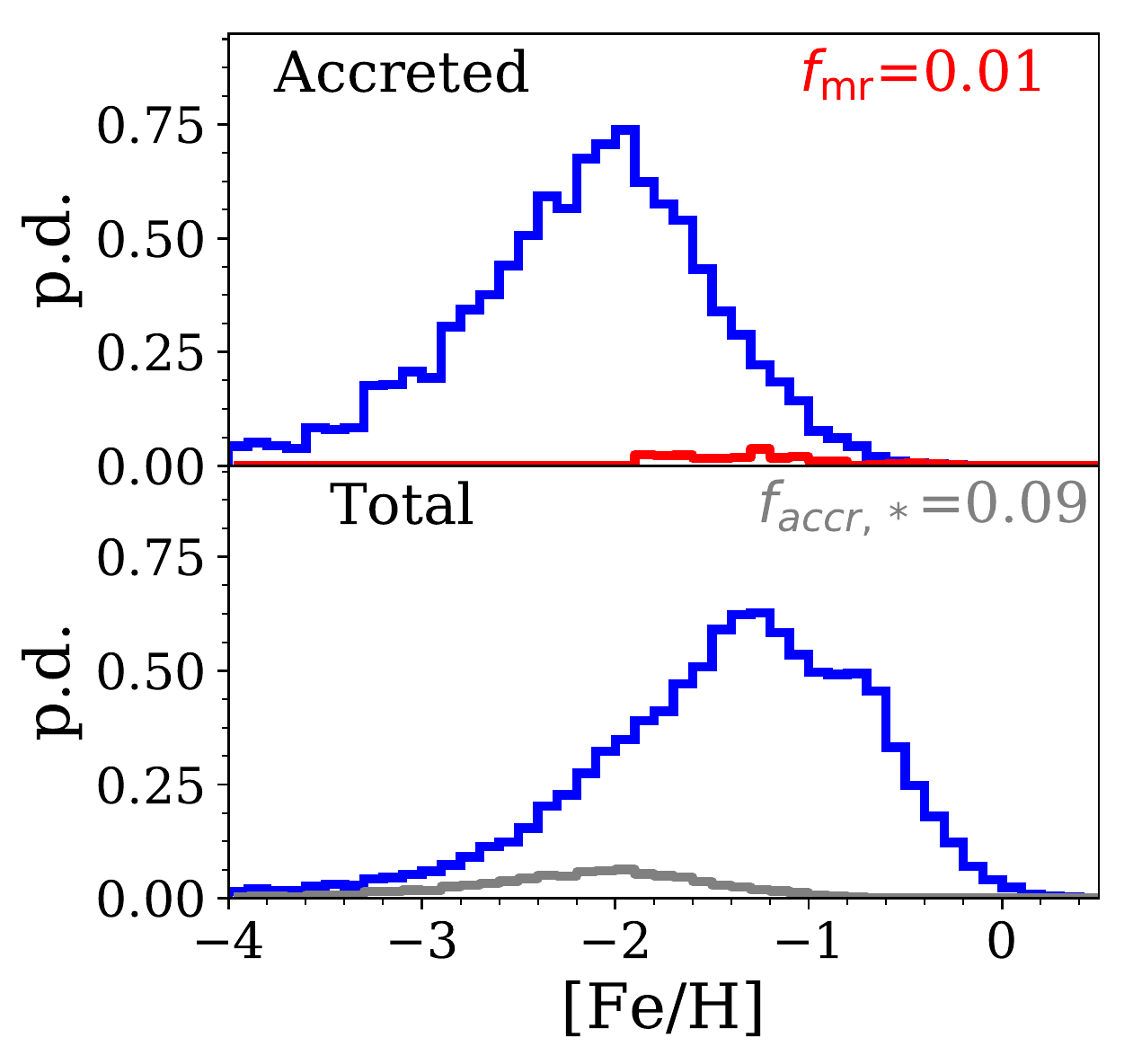}

		\caption{Similar to Fig.~\ref{feh_accrFig}, but for
                  satellite dwarfs with $\mathrm{r_{mr}/r_{mp}}$~<~0.65: histograms of
                  metallicity for stars that have been accreted
                  (upper figure) and for the total stellar
                  population (bottom figure).}
		\label{satfeh}
\end{figure}
\begin{figure*}
		\includegraphics[width=2\columnwidth]{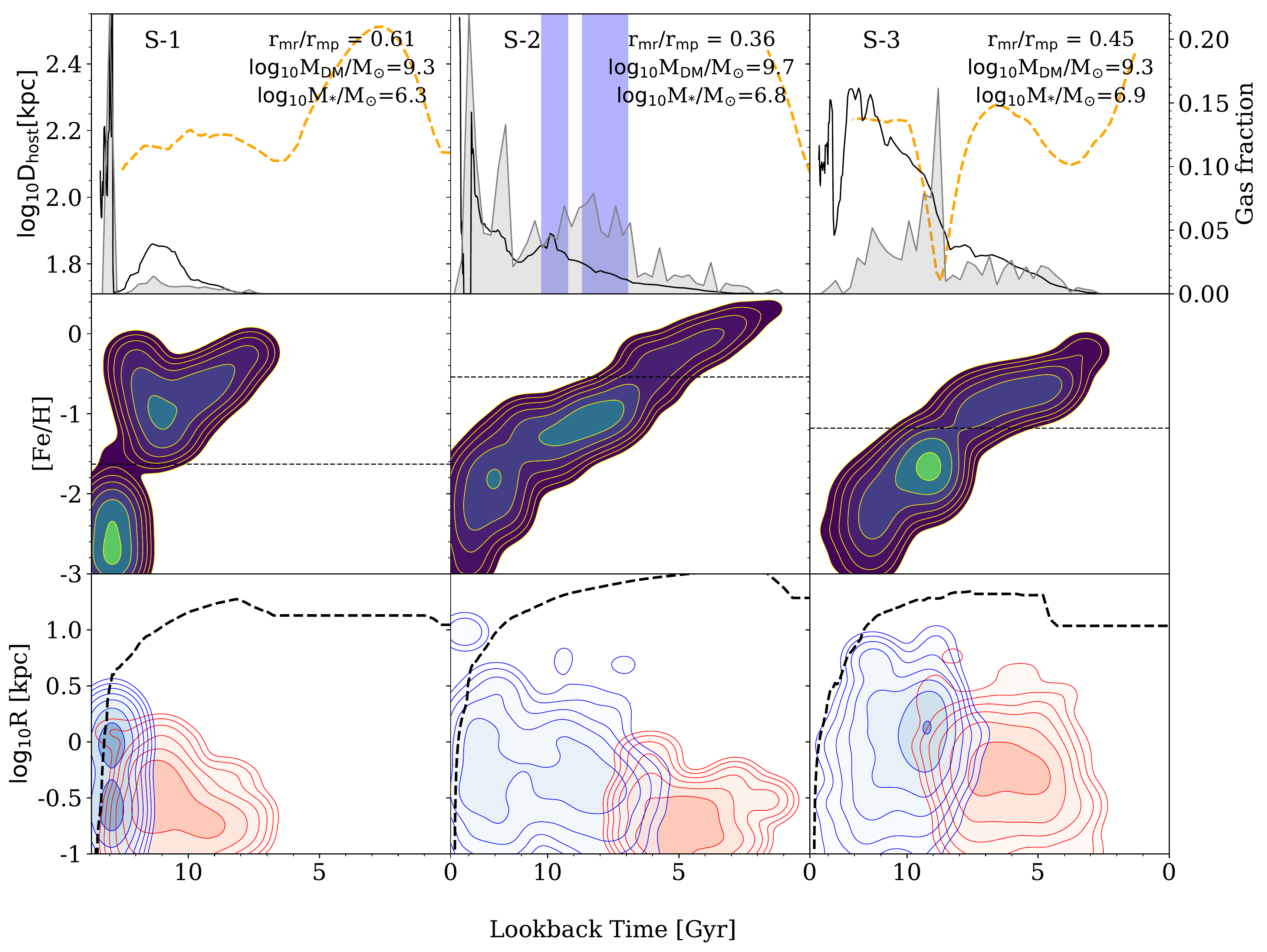}

		\caption{Properties of three example satellites
                 in order of increasing stellar mass. \textit{Top:} Grey shaded areas
                  show the SFH of the dwarfs in arbitrary units. The
                  black solid lines show the gas fraction of the
                  satellites (the bound gas mass divided by the dark
                  matter mass), with the corresponding scale displayed
                  on the right. The yellow dashed lines show 
                  distance to the host galaxy. For dwarf S-2, the blue bands show the times of interaction with cosmic filaments, as will be illustrated in Fig.~\ref{filament}.  \textit{Middle:} The metallicity-age distribution for the three galaxies. The dashed line shows the metallicity value at which our method splits the two populations. The probability density contours are geometrically spaced with a ratio of 2 and are consistent for the three dwarfs. \textit{Bottom:} The radii at which the the metal-rich (red) and the metal-poor (blue) particles form. The black dashed line is the evolution of the virial radius, which becomes the tidal radius after infall. The width of the smoothing kernel of the density estimate is consistent for the three galaxies across the middle and bottom panels.}

		\label{ramfigure}
\end{figure*}

\begin{figure*}
		\includegraphics[width=2\columnwidth]{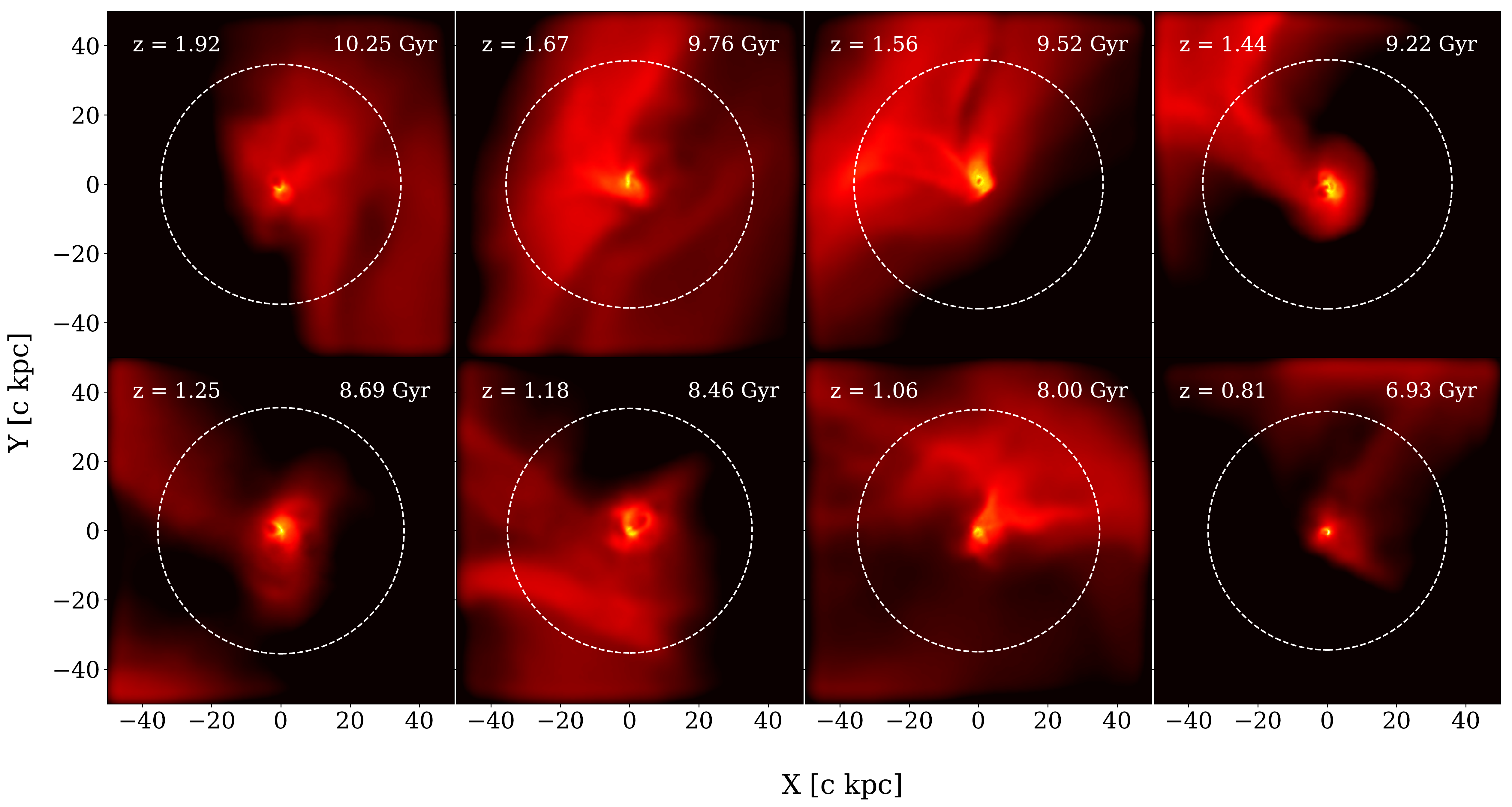}

		\caption{Gas density of dwarf
                  S-2 as it passes through two cosmic filaments (top and bottom panels, respectively),
                  which enhances star formation activity. The
                  redshift and lookback times are given in the upper
                  corners and the white circle represents the virial
                  radius. The high density regions are shown in yellow and the low density regions in black.} 

		\label{filament}
\end{figure*}

\begin{figure*}
		\includegraphics[width=1.8\columnwidth]{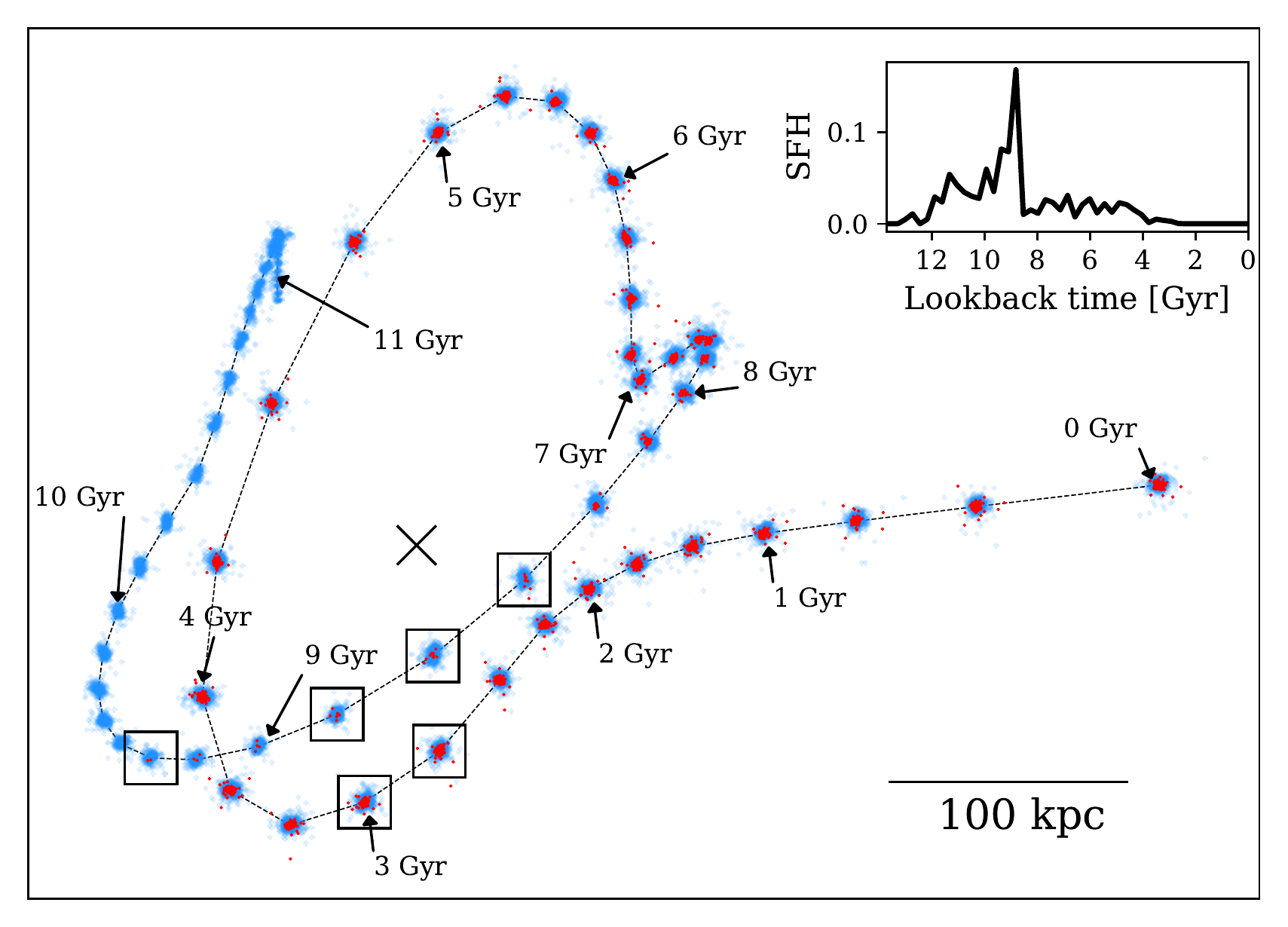}
		\includegraphics[width=1.8\columnwidth]{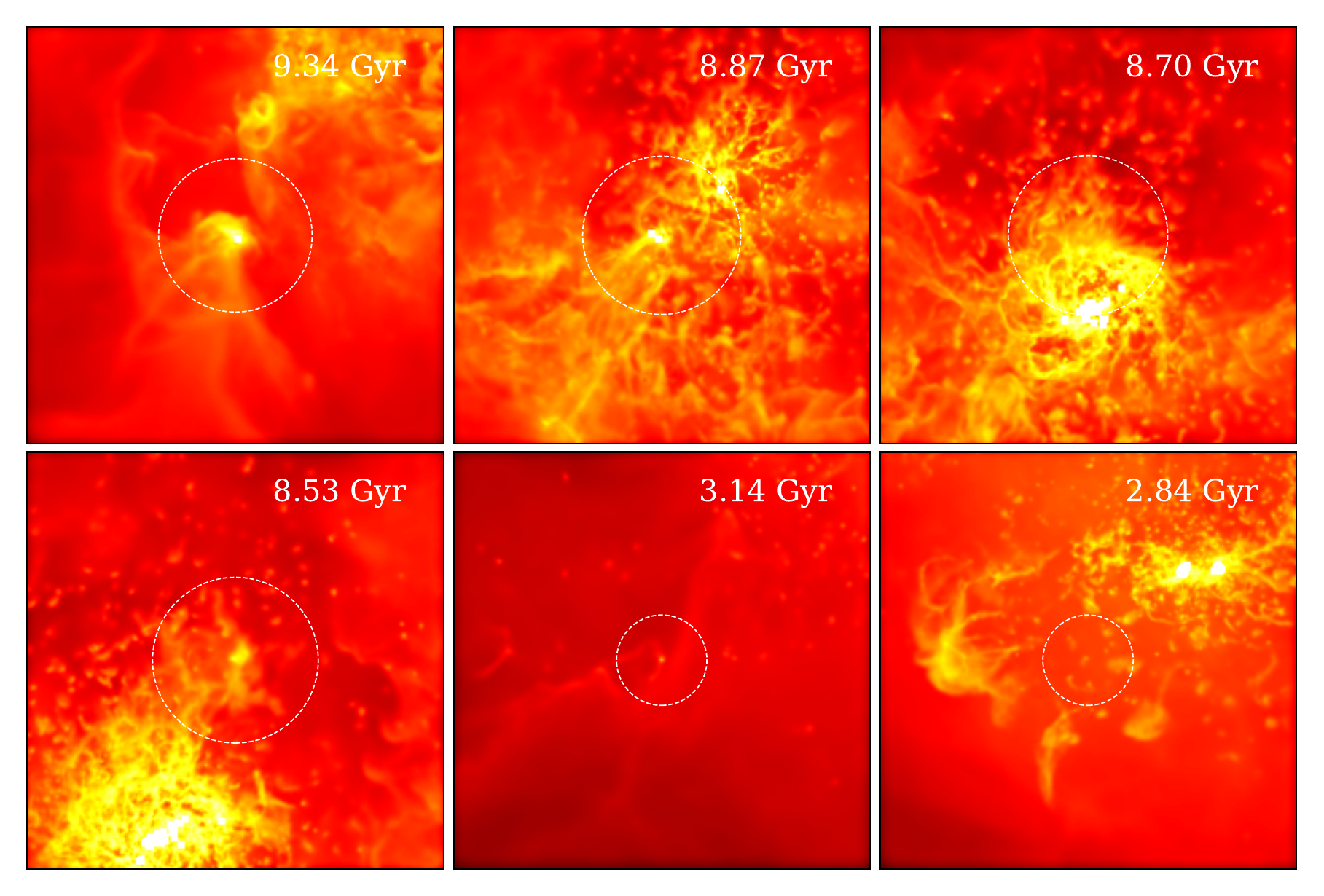}
        
                \caption{{\em Top panel}: The evolution of the
                  satellite (labelled S-3) since its first stars formed just before a
                  lookback time of 11~Gyr. The thin dotted line tracks
                  the orbit of the satellite in the reference frame of the host halo. The
                  black cross shows the position of the host halo. The squares mark important stages of the satellite's evolution. Between 9 and 8~Gyr in lookback time the satellite passes though the pericentre. At $\sim$3 Gyr it encounters a larger, more massive, satellite.  The inset at the top right displays the
                  star formation history. The abrupt changes in the
                  orbit between 8 and 6 Gyr, are caused by a
                  fluctuation in the centre of potential of the host
                  halo as it merges with a massive object. {\em Bottom panel:} the
                  evolution of the gas distribution during the six stages of the satellite's evolution, as shown with squares in the upper figure. Yellow shows high density and black shows low density regions. The circle
                  indicates the virial radius before infall ($\sim$9.5~Gyr) and the
                  tidal radius thereafter. The images show a cube of 50~kpc on a side.}

		\label{satmovie}
\end{figure*}

We now focus on the satellite dwarfs in the APOSTLE
simulations. Unlike isolated dwarfs, these may be subject to tidal and ram
pressure stripping, as well as other interactions with their hosts. As a
result, the origin of the two metallicity populations is different to isolated galaxies.

\subsection{Halo assembly in satellites}

Following our discussion of field dwarfs, we now investigate the role
of mergers in the formation of two metallicity populations
in satellites. Similarly to Fig.~\ref{accretionFig} and Fig.~\ref{feh_accrFig},
Fig.~\ref{sataccr} shows the fraction of accreted stars and
Fig.~\ref{satfeh} the distribution of metallicities for the accreted
and total stellar populations in the satellites.

A pattern similar to that in Fig.~\ref{accretionFig} may be seen,
although with larger scatter. The galaxies with greater spatial
segregation do indeed accrete larger fractions of their stars, though to a lesser extent than field dwarfs. As shown at the top of Fig.~\ref{satfeh}, approximately
1~per cent of the metal-rich population typically comes from outside
the galaxy and thus, as in field dwarfs, the metal-rich population is
predominantly formed {\em in-situ}. Unlike for field dwarfs, however, the
total metallicity distribution displays a more dominant metal-poor
population. For similar accretion fractions to those of field dwarfs,
this would suggest that either the mergers tend to occur later in
satellites, allowing less time for the metal-rich population to form,
or that processes other than mergers are at play that shut off star formation, resulting in a
smaller fraction of metal-rich stars.

We can conclude that mergers may play some role in the formation of
satellite dwarfs with two spatially segregated metallicity populations, yet lower accretion fractions and differences in the typical metallicity distribution compared to field dwarfs suggest that the precise mechanism may be more complex than in field galaxies.

We select the objects with $\mathrm{r_{mr}/r_{mp}}<0.65$ as our subsample of satellite dwarfs with two well spatially segregated populations. We will now examine the mechanisms by which the metal-rich population forms in these galaxies.

\subsection{Satellites with two metallicity populations and their environment}

We have seen some evidence that the satellites tend to have smaller fractions of metal-rich stars than the field dwarfs. The process of infall into the host halo limits the gas available to form the second population of stars and thus may be responsible for the
formation of a smaller metal-rich population. We now examine three examples of well spatially segregated satellite dwarfs in the context of their environment.

Fig.~\ref{ramfigure} shows the three example dwarfs in order of their stellar mass. The top panel shows the distance to the host galaxy (yellow line), the evolution of
the gas fraction (black) and the star formation history (grey). The middle panel displays the metallicity-age distribution for these galaxies and the lower panel shows the radii at which the metal-rich (red) and the metal-poor (blue) stars form.

\subsubsection{Two populations through mergers in satellites}

Dwarf S-1 has the lowest stellar mass of the two-population galaxies in our sample. At $\sim$13~Gyr in lookback time it merges with a gas-rich halo and forms its entire metal-poor population in a quick burst of star formation that consumes all of the available gas, expelling the rest from the halo through star formation feedback. This is evident in the evolution of the gas fraction seen for this dwarf in Fig.~\ref{ramfigure}. Following the burst, the galaxy smoothly accretes more gas that is used up over the next $\sim$5~Gyr in the formation of a metal-rich population. Note the peculiar shape of the metallicity-age distribution for this dwarf (middle panel). Some of the more metal-rich stars form first in this galaxy after the merger. This is because the first gas to return to the galaxy comes pre-enriched with metals expelled in the original outflow. The remainder of the accreted gas is slowly enriched as it sinks to the centre of the dwarf, with more metal-rich stars forming near the centre (bottom panel of Fig.~\ref{ramfigure}). Further accretion of gas in S-1 is prevented as the galaxy falls into its host halo with the first pericenter at $\sim7$~Gyr in lookback time.

\subsubsection{Ram pressure-induced star formation}
\label{ramsection}
 As can be seen in Fig.~\ref{ramfigure}, dwarf S-2 does not reach the pericentre of its orbit by $z$~=~0 and only falls in within the last 1~Gyr. At lookback times of $\sim13$~Gyr and $\sim 12$~Gyr it undergoes two accretion events, where its star formation peaks and the new stars are consequently formed with higher metallicities. Nonetheless, the metal-rich stars in this galaxy, as determined by our method, only begin forming at
$\sim$~9~Gyr, when the satellite passes through a gaseous cosmic
filament (blue band), triggering an increase in the star formation rate. At $\sim$~8~Gyr this galaxy passes through another gaseous filament. The metal-rich stars continue forming until the gas is depleted in this galaxy. Such
interactions have been previously noted by \citet{alecosmicweb} and
\citet{wright2018}. We show this passage through the cosmic filaments
in greater detail in Fig.~\ref{filament}. This displays the gas
distribution of the dwarf and its surroundings as it approaches the
cosmic filaments (the first filament is seen coming from the bottom right corner at 10.25~Gyr and the second filament in the bottom left corner at $\sim$8.69~Gyr) and passes through the overdense regions. Following a passage through each filament
the dwarf develops a tail of stripped gas. It is at these times
that S-2 undergoes the intense star formation activity that
enriches the interstellar medium and triggers the formation of metal-rich stars. Additionally, the stripping of the gas results in the preferential formation of these stars in the central regions, where the gas still remains bound (see bottom panel of Fig.~\ref{ramfigure}).

A small fraction of dwarfs form their two populations through interactions with other dwarf galaxies. The two dwarfs pass by each other without a merger, but with interacting dark matter haloes and gas, which results in the compression of the gas in the centre and its stripping in the outer regions. Note that these two mechanisms are also applicable to field dwarfs.

\subsubsection{Induced star formation by pericentric passage}

For dwarf S-3, the most intense episode of star formation activity is seen to
align with the first pericentric passage in Fig.~\ref{ramfigure}. In the metallicity-age distribution (middle panel) this is visible as a density peak associated with the metal-poor population, followed by a tail of metal-rich stars.

In Fig.~\ref{satmovie} we examine galaxy S-3 in particular detail. The top panel displays the orbit of the
satellite and the formation of its stars. The star formation history
is shown in the inset. The lower panel displays the gas
distribution, with the white circles representing the virial radius at times prior to infall and the tidal radius after infall\footnote{We define the tidal radius as the radius where the enclosed subhalo density is 3 times that of the host halo at that distance.}. 

This satellite falls into the host halo at $\sim$~9.34~Gyr. As can be
seen in the bottom panel, it still retains some amount of gas at
infall. Between 8.87-8.53 Gyr the satellite passes the pericentre of
its orbit and encounters the gaseous halo of its host, losing a
significant fraction of its gas through ram pressure stripping, as shown by
the tails in the gas distribution. However, during this time, star
formation in the satellite peaks and, by approximately 8~Gyr, the
first metal-rich stars begin to form. By $\sim$~3~Gyr the satellite
has turned most of its gas into stars and the remainder is further
stripped by an encounter with a larger subhalo, as seen in the bottom
right panel (at 2.84~Gyr). The abrupt changes in the satellite's orbit
between 8 and 6~Gyr in lookback time are a consequence of the
host galaxy undergoing a merger, resulting in a noticeable shift of 
its centre of potential whilst the merger is ongoing. We have verified
that the orbit of the satellite is, in fact, smooth in a different
reference frame.

The example of dwarf S-3 suggests a new
mechanism for the formation of two metallicity populations, whereby
near pericentre the satellite's innermost gas is compressed in the centre of the dwarf triggering a starburst. This gives rise to
significant enrichment of the interstellar medium over a short period
of time and to the formation of a metal-rich population at the
centre. Furthermore, due to ram pressure stripping, a large
portion of the remaining gas is subsequently lost and star formation is shut off.

\subsection{Summary}

We have identified three processes by which satellites can form their two metallicity populations. The first, as in the example of dwarf S-1, is consistent with a scenario we previously found for field dwarfs, with an extra caveat of an interaction with a host galaxy, where further growth of the metal-rich population is prevented by ram pressure stripping.

In galaxies S-2 and S-3 ram pressure and tidal interactions act to enhance the star formation activity and to enrich the interstellar medium, provided there is a pre-existing supply of gas. Furthermore, gas stripping allows only the innermost bound gas to remain to form stars, resulting in a high spatial segregation as well as a smaller fraction of metal-rich stars. The degree of spatial segregation
achieved through this mechanism is expected to be limited by the size of the metal-poor
population (in particular by whether or not it had been expanded by
earlier merger activity) and the amount of gas available for star
formation. Note that the Small and Large Magellanic Clouds have long been speculated to have episodes of enhanced star formation activity due to interactions with each other and the Milky Way \citep{lmcsmc}.

We note that merger activity or accretion is not uncommon in satellite
dwarfs (see Fig.~\ref{sataccr}). This plays a significant role in increasing the extent of the
metal-poor population, thus enhancing the spatial segregation,
although these events do not appear to be the cause of the formation
of the metal-rich population in the majority of our sample. 

Whether these scenarios are applicable to Local Group satellite galaxies is, of course, dependent on the realism of star formation histories exhibited by simulated dwarfs. For APOSTLE dwarfs, this has been recently investigated by \citet{digby}, who find general agreement with Local Group galaxies for which the photometry reaches the main sequence turnoff. 

The efficiency of gas stripping when a satellite falls into the halo of its host is one of the key aspects that determines whether it may continue to form stars. In Appendix~\ref{AppendixB} we show the mass in H\textsc{I} to light ratio for our dwarfs as a function of distance to the nearest host galaxy at $z=0$. These values agree well with the current measurements. In light of this, the scenarios described above are certainly plausible in Milky Way and Andromeda satellites.

\section{Mass dependence of the formation mechanisms}
\label{massdepsec}

\begin{figure}
		\includegraphics[width=\columnwidth]{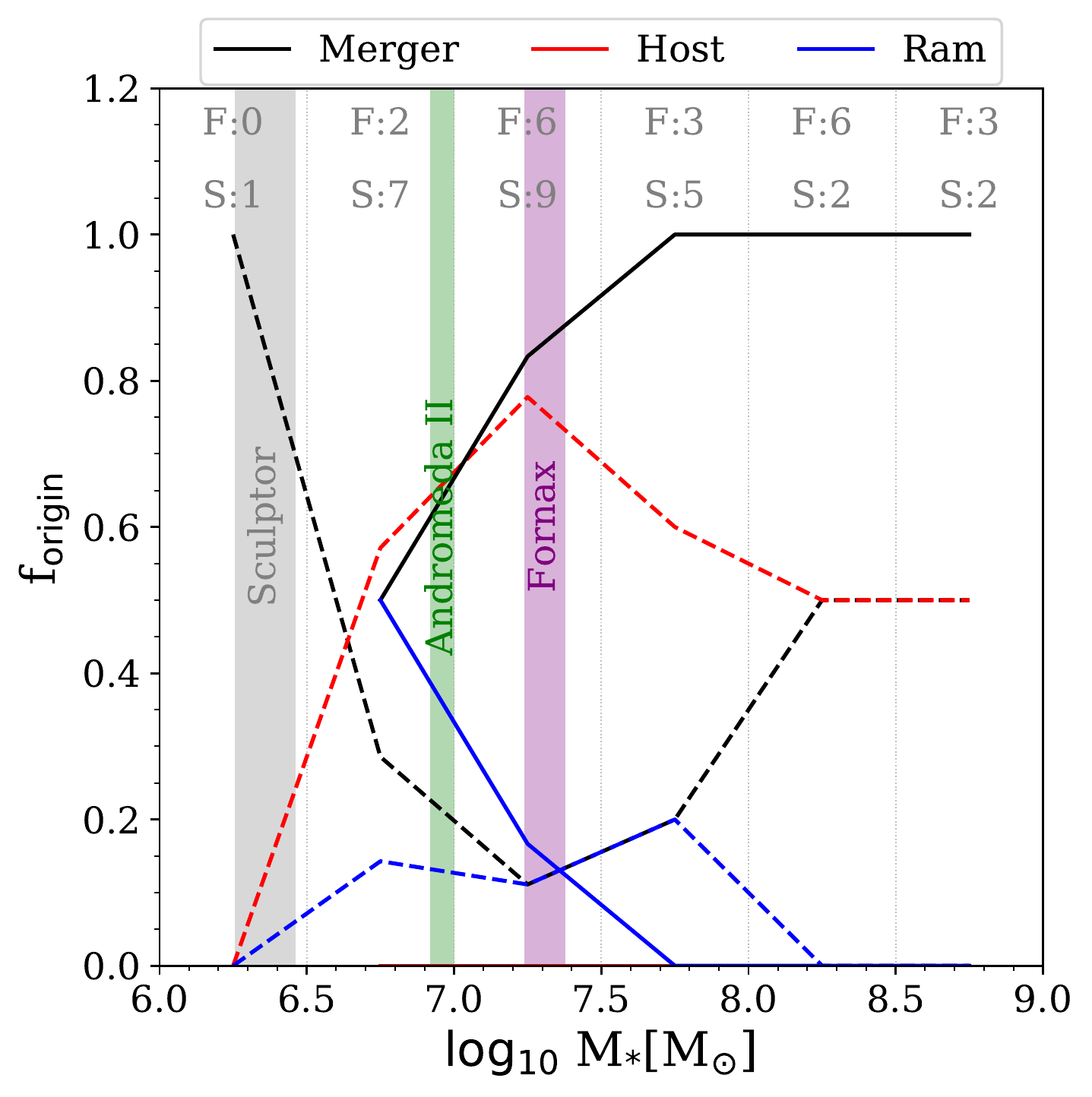}

		\caption{ The fraction of field (solid lines) and satellite (dashed lines) dwarfs forming through each mechanism: mergers (black), interaction with a host galaxy (red, for satellites only) and ram pressure and tidal interactions with cosmic filaments or other galaxies (blue). Bins in stellar mass are separated with grey dashed lines. At the top, we show the number of field and satellite dwarfs within each bin. The stellar masses (assuming stellar mass-to-light ratio of 1) for Sculptor, Fornax, and Andromeda~II are shown with coloured bands. }

		\label{massdep}
\end{figure}

In this Section we explore how the formation mechanisms
described for field and satellite dwarfs with two spatially segregated populations operate for different bins in stellar mass. For each dwarf in this sample we determine the primary cause of formation of the metal-rich population through analysis of the metallicity-age relation, as well as merger and environmental histories. In particular, we note the event responsible for sufficient gas enrichment that results in the formation of metal-rich stars. The results are shown in Fig.~\ref{massdep}, where we explore 6 stellar mass bins, $\log_{10}$M$_*$/M$_{\odot}$~=~[6, 6.5, 7, 7.5, 8, 8.5, 9], highlighted with grey dashed lines.

For field dwarfs with $\mathrm{r_{mr}/r_{mp}}$~<~0.65  (solid lines) we find that $\sim$90 per cent form their two populations through mergers (black), with the remainder forming through interactions with cosmic filaments and other dwarfs (blue). We find that the probability of formation through mergers steadily increases with stellar mass, as expected for more massive haloes. 

We find that $\sim$60 per cent of the satellites with $\mathrm{r_{mr}/r_{mp}}<0.65$ (dashed lines) form their metal-rich population by passing through pericenter (red), $\sim$30 per cent form through mergers (red) and $\sim$10 per cent through interaction with filaments or nearby galaxies (blue). The pericentre mechanism for satellites dominates at all stellar masses above  $\log_{10}$M$_*$/M$_{\odot}$~=~6.5, remaining consistently at above $\sim$50 per cent. Above $\log_{10}$M$_*$/M$_{\odot}$~=~7.5 the fraction formed through mergers increases from $\sim$10 per cent to 50 per cent. Below $\log_{10}$M$_*$/M$_{\odot}$~=~6.5 only S-1 is present in our sample, which formed through a merger. There is some evidence for an increase in the formation of a metal-rich population through mergers below $\log_{10}$M$_*$/M$_{\odot}$~=~7. It is difficult to say whether in this mass range the merger scenario would dominate, given the sample size, although these lower mass objects are certainly less likely to maintain sufficient amounts of gas to form a metal-rich population at pericenter. 

The grey, green and purple bands show the mass ranges for Sculptor, Andromeda~II and Fornax, respectively (assuming stellar mass-to-light ratio equal to 1) \citep{mcconnachieCensus}. As satellites, Andromeda~II and Fornax are most likely to have formed their metal-rich populations during their pericentric passages; however this scenario is dependent on these dwarfs passing though pericentre while retaining a sufficient supply of star-forming gas. Otherwise, mergers and interactions with other subhalos or filaments are likely to play the main role. The analyses of dwarf star formation histories and proper motions, in light of results from Gaia \citep{gaiadwarf}, should shed light on the applicability of the pericenter scenario to individual dwarfs.

\section{Properties of two-metallicity population dwarfs}
\label{signaturessection}

\begin{figure}
        \centering
		\includegraphics[width=\columnwidth]{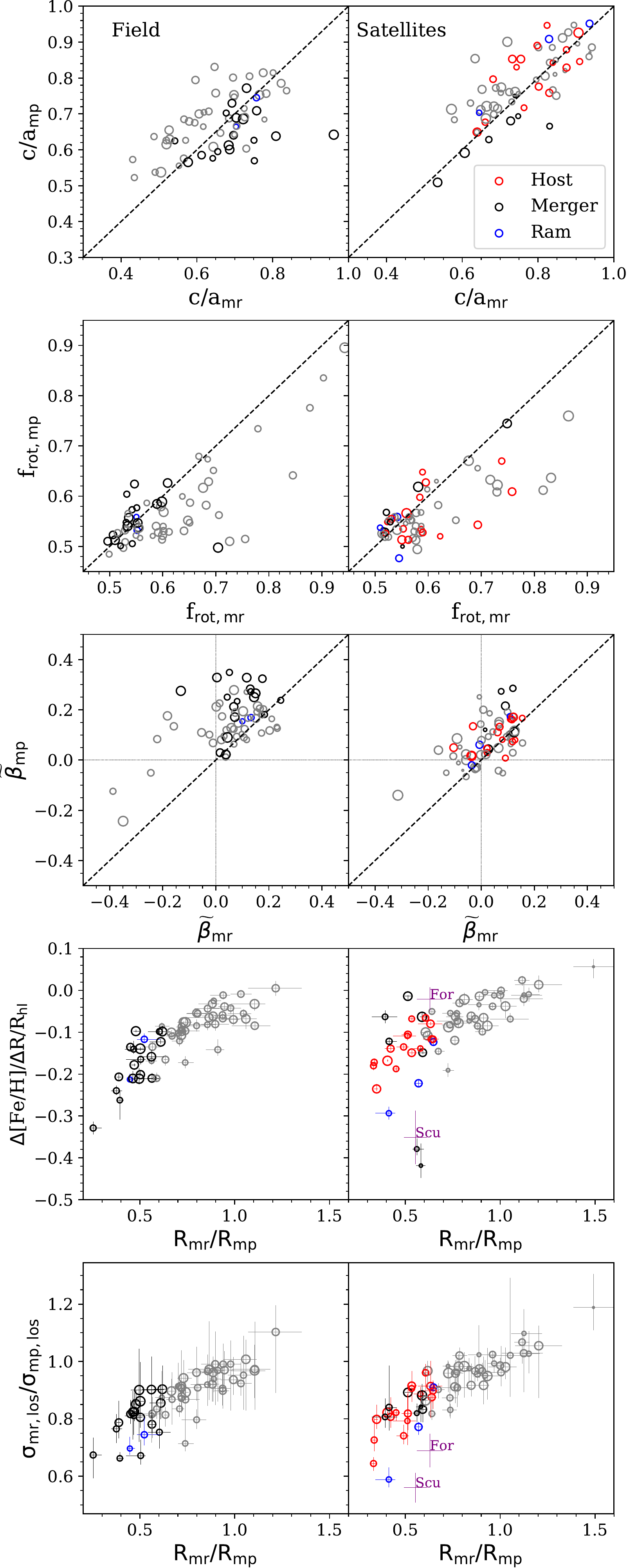}

		\caption{Properties of individual
                  subpopulations. Colours denote formation mechanisms (see legend). Symbol sizes indicate the logarithm of stellar mass. The black dashed lines
                  are one-to-one relations. \textit{First row:} The
                  sphericity of the two populations, $c/a$. \textit{Second
                    row:} The fraction of stellar particles in ordered rotation, $\mathrm{f_{rot}}$. \textit{Third row:} 
                  Velocity anisotropy, $\widetilde{\beta}$. \textit{Fourth row:} The metallicity
                  gradient, $\mathrm{\Delta[Fe/H]/\Delta R/R_{hl}}$, as a function of the ratio of the projected
                  half-mass radii, $\mathrm{R_{mr}/R_{mp}}$. \textit{Fifth row:} The ratio
                  of the line-of-sight velocity dispersions as a function of $\mathrm{R_{mr}/R_{mp}}$.  }

		\label{sphkap}
\end{figure}

In this Section we explore properties of two-population satellites and
field dwarfs. Of particular interest are the ellipticities of the
individual populations, the significance of rotation, velocity
anisotropy, metallicity gradients and the velocity dispersion ratio of the two populations. We investigate whether these
properties contain information about the formation paths of these 
galaxies. We show these properties in Fig.~\ref{sphkap}, which displays satellite and field dwarfs with $\mathrm{r_{mr}/r_{mp}}$~<~0.65 with coloured symbols, with the rest of dwarfs that show bimodality in their metallicity distributions shown in grey. The size of the symbols reflects the logarithm of the stellar mass of the dwarfs.

We define sphericity as the ratio, $s = c/a$, where $c^2$ and $a^2$ are
the eigenvalues of the reduced inertia tensor corresponding to the minor and
major axes of a 3D ellipsoid \citep{sphericity}. An isotropic distribution would have $s = 1$. The
upper panels of Fig.~\ref{sphkap} show the individual sphericities of
the metal-rich (horizontal axis) and the metal-poor populations (vertical
axis). Field dwarfs are shown in the left-hand panel and
satellites in the right. The black circles represent dwarfs with two spatially segregated metallicity populations, where the formation of the metal-rich-population has occurred as a result of a merger, while
the red circles show satellites where the metal-rich population has formed as a result of the first pericentric passage. The blue circles show objects where the metal-rich population is the result of a galaxy's interaction with a cosmic filament or a nearby galaxy. It is clear that satellites
tend, on average, to be more spherical than field
galaxies. Additionally, the subset of satellites whose star formation
peaks near first pericentre also tends to be more spherical than
the rest of the sample.
  The field dwarfs with high spatial segregation tend
to have more spherical metal-rich populations. This could be a
reflection of the effects that the mergers responsible for the creation of a metal-rich population have on the shape of these galaxies.

We characterize the rotation of individual subpopulations by the
fraction of stars within that subpopulation that are rotating in the same direction. In the second row of
Fig.~\ref{sphkap} we show these fractions for the metal-rich, $\mathrm{f_{mr}}$, and the metal-poor, $\mathrm{f_{mp}}$, populations. The
metal-rich populations tend to exhibit stronger rotation than the
metal-poor populations, in both satellites and field dwarfs.  Field
dwarfs with high spatial segregation appear closer to the one-to-one
relation, whilst satellites with star formation peaking near
pericentre show a small bias towards more rapidly rotating metal-rich
populations. Evidence of rotation has been found in real
two-population Local Group dwarfs
\citep{battagliaSculptor,andromedaii,delpinofornaxrotation}.

We define the orbital anisotropy as
$\beta = 1 - \sigma_t^2/(2\sigma_r^2)$, where $\sigma_t$ and
$\sigma_r$ are the tangential and radial velocity dispersion
components respectively. In the third row of Fig.~\ref{sphkap} we plot
the anisotropy, normalised to lie between 1 (radial) and -1
(tangential), $\widetilde{\beta} = \beta/(2-\beta)$
\citep{gravsphere}. These values are averaged in bins of equal
particle number. The dotted lines indicate isotropy. It is clear that
stars in field dwarfs have preferentially radially-biased motions that
are stronger in the metal-poor population. This behaviour is less
extreme in satellites perhaps reflecting preferential stripping of
radially-biased orbits.

The fourth row of Fig.~\ref{sphkap} shows the projected metallicity
gradients of our satellite and field dwarfs as a function of $\mathrm{R_{mr}/R_{mp}}$. The uncertainties were derived by
calculating these quantities along 1536 evenly distributed lines of
sight \citep{healpix}. The metallicity gradients are characterized by
the slope of the least-squares fit to [Fe/H] and R/R$_{\mathrm{hl}}$,
where R$_\mathrm{hl}$ is the projected half-light radius. In this fit we exclude the extremely metal-poor stellar particles ([Fe/H] < -4) as these are unreliable given the mass resolution of our simulations \citep{eagleschaye}. We also only include the stars within 5R$_{\mathrm{hl}}$, which excludes the outer halo stars that would be difficult to differentiate from background in observations.
Galaxies with strong spatial segregation exhibit particularly steep gradients, in some cases, more than twice as steep as the typical value for the sample ($\Delta$[Fe/H]/$\Delta$(R/R$_{\mathrm{hl}})\sim-0.1$). This is consistent with the metal-poor halo
assembly through early mergers seen in the simulations of
\citet{jablonka}. The purple symbols show the metallicity
gradients in Sculptor and Fornax, respectively, obtained from 
\citet{kirbygradient}, where we take the $\mathrm{\Delta}$[Fe/H]/$\Delta$r values and multiply by the half-light radii from \citet{mcconnachieCensus}, carrying the errors, for better comparison with what we measure in our simulations. These observations are consistent with the
steep metallicity gradients present in satellites in our simulations
with well spatially segregated populations. In particular, note that Sculptor, overlaps within 1$\sigma$ with two satellite dwarfs that has formed via a merger (black points). One of these galaxies is dwarf S-1 and anther follows a similar formation history.

Finally, the bottom panel of Fig.~\ref{sphkap} shows the ratio of the
line-of-sight velocity dispersions of the two populations,
$\sigma_{\mathrm{mr,los}}/\sigma_{\mathrm{mp,los}}$, plotted against
the ratio of the projected half-mass radii. These quantities are of
particular interest in the application of the \citet{walkerPenarrubia}
method for the determination of the inner dark matter density
slopes. The two quantities follow a very clear relation. This is not
surprising, as the velocity dispersion is expected to scale with enclosed mass. Satellites appear to have a less significant kinematic difference between the two populations than field dwarfs. \citet{sales} show that this may be due to the effects of tidal stripping. We have investigated this and found only a weak trend, which may be due to star formation at pericenter, which \citet{sales} do not include in their analysis. The
values for Sculptor and Fornax, taken from
\citet{walkerPenarrubia}, are plotted in the figure. These values lie
well below our relation even after taking into account the effects of
projection. The origin of this discrepancy is unclear and might be due
to the differences in the way in which the two metallcity populations
are defined in the simulations and in the data; in particular, metallicity mixing between the two populations may have an effect. Additionally, the larger typical sizes of our sample of dwarfs compared to real dwarfs may contribute to this discrepancy, in particular by increasing the velocity dispersion of the metal-rich population, although the effect of the larger overall sizes on $\mathrm{R_{mr}}/\mathrm{R_{mp}}$ is not entirely clear.

From these considerations we conclude that the spatial and kinematic
properties of individual subpopulations do not provide sufficient
information on the formation paths of the two subpopulations; many of
the observed differences are primarily due to environmental effects
(satellites or isolated dwarfs) and seem to be independent of stellar mass. Nonetheless, we have seen that
satellites tend to have a smaller number of metal-rich stars relative
to metal-poor stars than field dwarfs. This is
almost certainly due to the smaller gas supply available to form the
metal-rich stars due to tidal and ram pressure stripping of gas as the
satellite falls in. In this case, the constraints on the stellar ages, metallicities and the satellite orbit should allow one to distinguish between the merger or pericentric passage origin of the metal-rich population.

\begin{figure*}
        \centering
		\includegraphics[width=2\columnwidth]{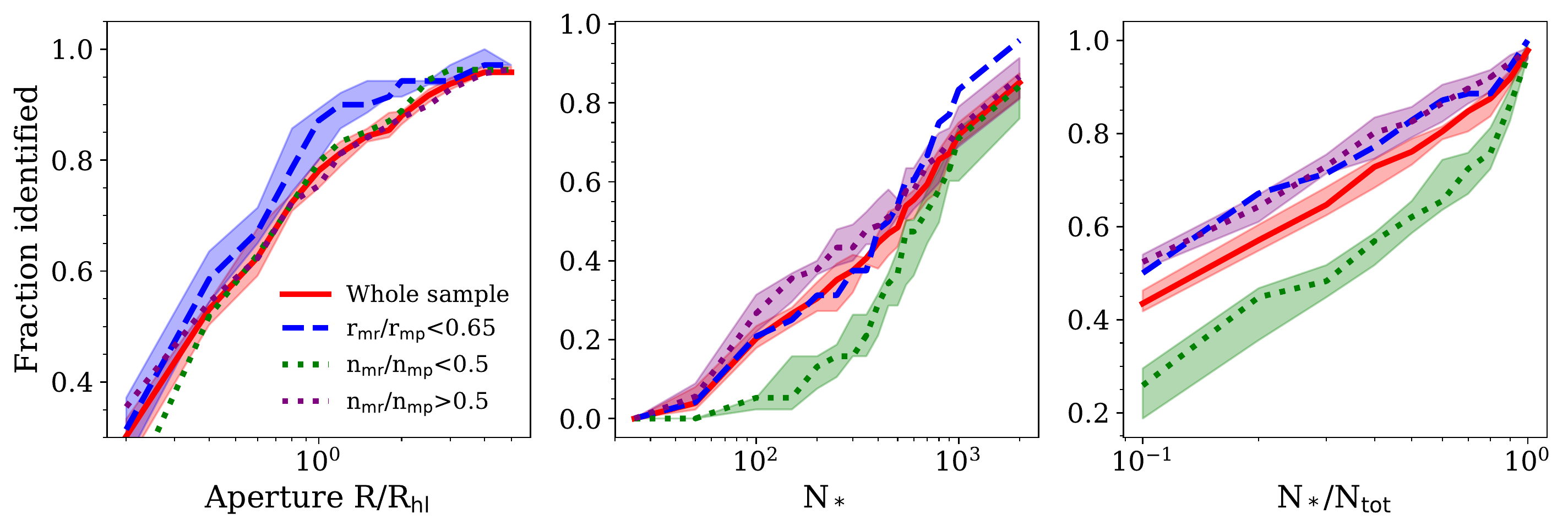}

		\caption{The fraction of two-population dwarfs detected when varying three key quantities. \textit{Left:} The aperture within which the stellar sample is detected as a function of the projected half-light radius. \textit{Middle:} The number of stars used within each galaxy.  \textit{Right:} Fraction of the total number of stars. The colours show different subsamples of dwarfs, with the whole sample of two-population dwarfs in red; those that are well spatially-segregated in blue; dwarfs with a dominating metal-rich population in purple and dwarfs with a dominating metal-poor population in green. The shaded bands show the 16$^{\mathrm{th}}$ and 84$^{\mathrm{th}}$ percentiles from varying lines of sight. We only show the shaded bands for those subsamples of dwarfs that deviate significantly from the results for the whole sample.}

		\label{convergence}
\end{figure*}

\section{Metallicity bimodality in Local Group dwarfs}

\label{observations}

In this work we have identified a sample of dwarf galaxies within the APOSTLE suite of simulations that exhibit a bimodality in their [Fe/H] distribution, which we define as an indicator of the presence of two metallicity populations. These galaxies make up nearly half of our overall sample in the stellar mass range 3$\times10^6$ - 1$\times10^9$~M$_{\odot}$. Whilst signs of bimodality have been seen in the metallicity distributions of Sculptor \citep{kirbysculptor} and Fornax \citep{battagliaFornax}, the shapes of the distributions also tend to change with different or updated samples of stars \citep{kirbyrelation}. In fact, the work of \citet{walkersculptor}, with data for over 1000 stars in both Sculptor and Fornax, shows no trace of two metallicty peaks\footnote{Note that \citet{walkersculptor} use spectroscopically determined magnesium and iron line indices as indicators of [Fe/H]. These authors note that the indices are better used as measures of \textit{relative} metallicity.}. We thus aim to establish to what extent does the sampling of stars and the spatial extent of the observed region affect the effectiveness of our method, described in Section~\ref{defining}, in determining the number of populations in a given dwarf.

We begin by considering our entire sample of two-population dwarfs. For each of these dwarfs we select 192 evenly spaced lines of sight with the method of \citet{healpix} and change the size of the ``aperture" in units of the projected half-light radius, ranging between 0.2R$_{\mathrm{hl}}$ and 5R$_{\mathrm{hl}}$. We select the stars contained within the aperture and apply our method for determining the number of populations. The results are shown with the red line in the left panel of Fig.~\ref{convergence}, where we show the fraction of two-population galaxies identified correctly as a function of radius. Note that this requires all stars to be observed within this radius. The shaded regions represent the 1$\sigma$ scatter from different lines of sight. It can be seen that the stars extending out to at least 2R$_{\mathrm{hl}}$ are required to correctly identify all galaxies. Within the half-light radius $\sim$80 per cent of systems are identified. This statistic is improved for dwarfs with a significant spatial segregation between the two populations (blue line), where the data within the half-light radius is sufficient to identify two peaks. Note that whilst the two peaks may be found, the spatial segregation may not be apparent if the metal-poor population is particularly extended. The galaxies with a dominant metal-rich (purple) and a dominant metal-poor (green) populations do not significantly differ from the trend for all galaxies.

In the middle panel of Fig.~\ref{convergence} we show the fraction of correctly identified galaxies as a function of the number of stellar particles within the sample. The shaded bands indicate the scatter due to particle sampling. Here we only include dwarfs with at least 2000 stellar particles. It can be seen that even with a sample of a 1000 stars only 70 per cent of the dwarfs show bimodality. For a typical sample of $\sim$300 stars only 40 per cent show two metallicity peaks. The fraction is smaller for dwarfs with a dominant metal-poor population. This is characteristic of satellites that form their metal-rich population through the pericentric passage route. This is also evident in the rightmost panel, where we show the fraction of the total number of stars within the galaxy. A full sample is required to ensure 100 per cent detection of two peaks. From the leftmost panel, we can see that this requires data from within $\sim$2-3R$_{\mathrm{hl}}$. 

From the sample of \citet{mcconnachieCensus}, there are collectively 25 Milky Way and Andromeda satellites that fall within the magnitude range explored in this paper. If, similarly to our sample, 40 per cent of these have bimodality in their metallicity distribution ($\sim$ 10 galaxies), then provided a sample of spectroscopically determined metallicities and a typical sample size of $\sim$300 stars, we expect that 2-5 of those will be found to exhibit bimodality. 

With the above in mind, we examine the data provided by \citet{kirbyrelation} for signs of bimodality in the metallicity distributions. This is described in Appendix~\ref{AppendixC}, where our method only identifies two populations in Sculptor, with a metallicity cut between the two consistent with \citet{battagliaSculptor}. The spatial extent of these data however does not allow inferences to be made about the spatial segregation of the two populations. 

We conclude that whilst the sample sizes currently observed may be sufficient for our method to identify bimodality, further spectroscopic measurements are needed, in particular for the metal-poor population in the outer regions of dwarfs, although obtaining these may be extremely challenging due to increasing background contamination.

\section{Conclusions}

\label{conclusions}

A number of dwarf galaxies of the Local Group contain two spatially
and kinematically distinct stellar populations. The origin of this
phenomenon is not fully understood although several scenarios have
been proposed in the literature. In particular mergers are thought to
play an important role \citep{Lokasrotation,Starkenburg,alejandro-mergers,fouquet}. This scenario has some backing from
observations of substructure and rotation in dwarfs like Fornax and
Andromeda II. 

In this work we have examined 142 field and 108
satellite dwarf galaxies from the APOSTLE suite of cosmological
hydrodynamics simulations. We find that 43 per cent of field dwarfs and 53
per cent of satellite dwarfs show bimodality in their [Fe/H]
distributions and among those $\sim$30 per cent are well spatially segregated.

We first consider field dwarf galaxies. We find that their stellar
accretion fraction is directly related to the degree of spatial
segregation between the two metallicity populations. We find evidence
that this is primarily due to the metal-poor stars migrating to larger orbital
radii as a result of a merger. The
metal-rich population is subsequently formed {\em in-situ}. Among the field dwarfs with two well-segregated metallicity populations, $\sim$~90 per cent form their metal-rich populations by this mechanism.

In addition to the merger scenario for the formation of two
population dwarfs, we identified a formation mechanism that is specific
to satellite galaxies. As the satellite falls into its host halo, ram pressure may compress the gas at
the centre of the satellite whilst simultaneously stripping gas from
the outer regions. As a result, a new population of stars forms at the centre
of the satellite. Of the satellites with two well spatially segregated populations $\sim$~60 per cent form their metal-rich population through this mechanism.

A related process which occurs in both satellites and field dwarfs is
interaction with a gaseous cosmic filament. When a galaxy crosses a
filament, ram pressure may
trigger star formation activity, whilst simultaneously stripping
the outer gas. Nonetheless, we find that this mechanism is responsible for the formation of metal-rich stars in only $\sim$10 per cent of galaxies with two well segregated populations.

We also investigate the properties of the individual metallicity
subpopulations, specifically rotation, sphericity and velocity
anisotropy. In general, we find that both populations tend to have
higher sphericities in satellites than in field dwarfs. This is
consistent with previous work on the effects of tidal stripping
\citep{spherical_barber}. In field dwarfs the metal-poor population is
typically rounder than the metal-rich population; stars in both
populations tend to have radially-biased orbits, more so in the
metal-poor population. In satellites this bias is smaller, due to the
preferential stripping of stars in radial orbits.

The presence of two metallicity populations results in metallicity
gradients similar to those in
two-population Local Group dwarfs, particularly Sculptor and Fornax;
these resemble galaxies in our sample that have undergone mergers. On
the other hand, the ratio of the line-of-sight velocity dispersions of
the two populations in Sculptor and Fornax appear inconsistent with
the ratios in our sample. The origin of this discrepancy is unclear
but it could be influenced by differences in the definitions of the two
metallicity populations. 

We investigated whether the spatial and kinematic information on the
individual populations could help identify their formation mechanism,
particularly in satellite dwarfs. We found that the available
information is insufficient for this purpose. Nonetheless, measurements of stellar metallicities and constraints on dwarf orbits and star formation histories will provide important clues on their likely origin.

We have explored how many Local Group satellites, in the magnitude range similar to that explored in this paper, are likely to be identified as having a bimodal metallicity distribution given observational limitations. Our results suggest that approximately 2-5 objects should be detectable, given spectroscopic samples of $\sim$300 stars. We use our method to identify two populations in Sculptor dwarf spheroidal, consistent with previous works. We point out that data with a large spatial extent, (> 2R$_\mathrm{hl}$), which contains over 1000 stars is required for $\sim$70 per~cent detection and is particularly important for identifying galaxies with dominant metal-poor populations, such as those that have formed their metal-rich population by passing through pericentre.

\section*{Acknowledgements}
This work benefited from the use of \textsc{scikit-learn} \citep{scikit-learn} and
\textsc{py-sphviewer} \citep{sphviewer}. The authors thank Till Sawala, Louis Strigari and Evan Kirby for useful communications. We would like to thank an anonymous referee for detailed and constructive
comments that led to significant improvements in this paper. This work was supported by
the Science and Technology Facilities Council (STFC) consolidated
grant ST/P000541/1. AG acknowledges an STFC studentship grant
ST/N50404X/1. CSF acknowledges support by the European Research
Council (ERC) through Advanced Investigator grant DMIDAS (GA 786910). JFN acknowledges the hospitality of the KITP at UC Santa Barbara and of the Aspen Center for Physics. This research was supported in part by the National Science Foundation under Grants No. NSF PHY-1748958 and PHY-1607611. KO received support from VICI grant 016.130.338 of the Netherlands Foundation for Scientific Research (NWO). AF acknowledges support by a European Union COFUND/Durham Junior Research Fellowship (under EU grant agreement no. 609412). This work used the DiRAC Data Centric system at Durham University,
operated by the Institute for Computational Cosmology on behalf of the
STFC DiRAC HPC Facility (\url{www.dirac.ac.uk}). This equipment was
funded by BIS National E-infrastructure capital grant ST/K00042X/1,
STFC capital grant ST/H008519/1, and STFC DiRAC Operations grant
ST/K003267/1 and Durham University. DiRAC is part of the National
E-Infrastructure. This work used the UK Research Data Facility
(http://www.archer.ac.uk/documentation/rdf-guide)

\bibliographystyle{mnras} 
\bibliography{popbib}

\appendix

\section{Bimodality in the metallicity distribution}

\begin{figure}
		\centering
		\includegraphics[width=\columnwidth]{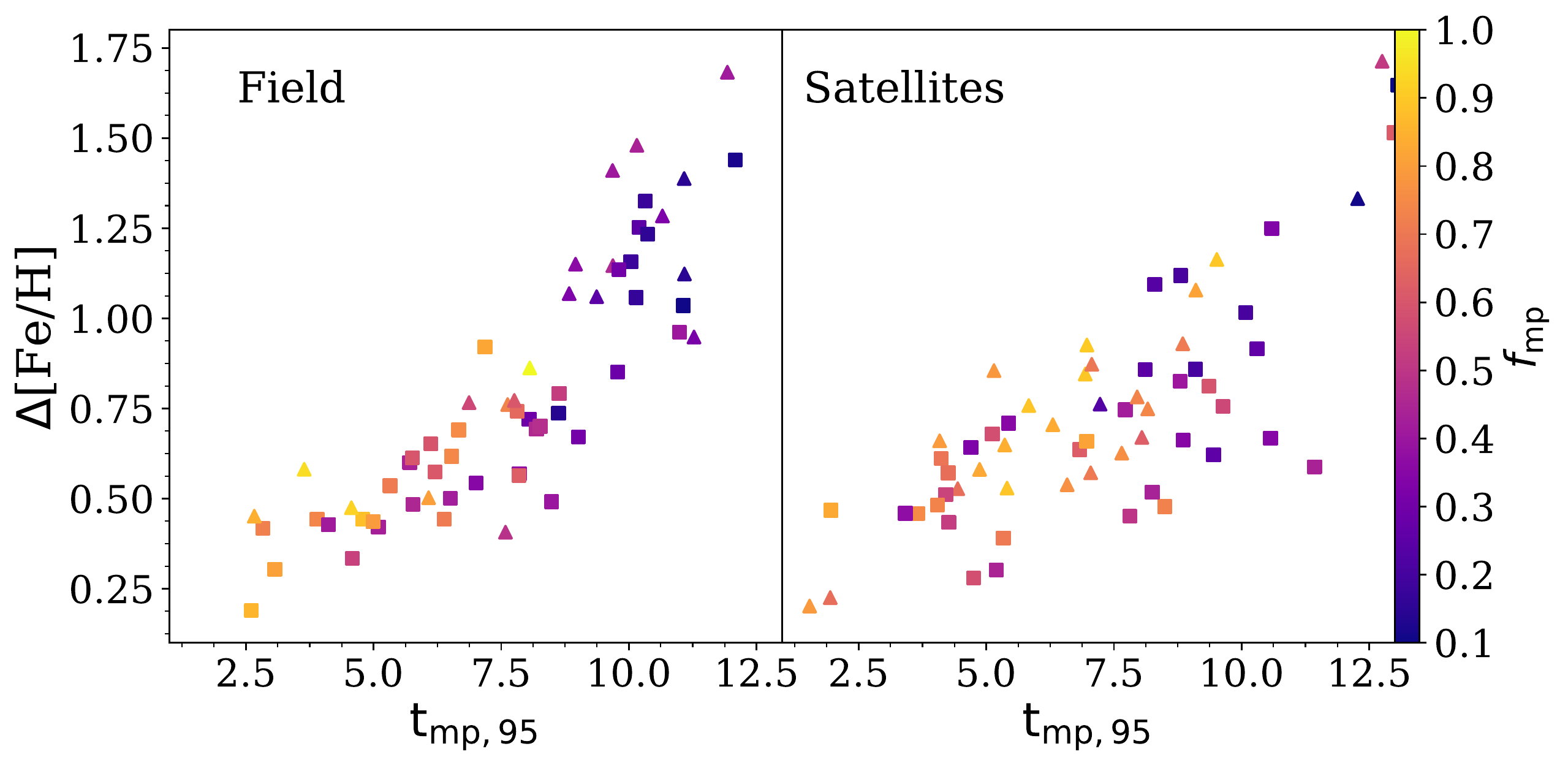}

		\caption{The `gap' between the peak
                  metallicities of the metal-rich and metal-poor
                  populations as a function of the lookback time to
                  when the metal-poor population stopped forming,
                  $\mathrm{t_{mp,95}}$. The points are coloured by the
                  fraction of metal-poor stellar mass. Field galaxies are shown on the left and satellites on the right. The galaxies
                  that have a particularly large spatial segregation
                  are shown with triangles. It is clear that these are
                  not dependent on the size of the metallicity peak
                  separation.  }

		\label{enrichment1}
\end{figure}

We discuss the origin of the bimodality in the metallicity distribution that allows our method to identify two metallicity populations. In particular, we focus on the creation of a
`gap', or the separation, between the two metallicity peaks and of the
`dip' in the distribution (i.e the minimum between the two metallicity
peaks). We note that the `dip' can be a feature of the Gaussian
mixture fit, rather than of the metallicity distribution itself. In
fact, the region where the metallicity cut is placed can be rather flat in certain cases. This region between the two metallicity peaks
does, however, play a role in determining the optimal number of
Gaussian mixtures in the fit to the distribution and therefore our ability to
identify two peaks.

\subsection{The gap between the two metallicity peaks and spatial segregation}
\label{gap}

Given a supply of gas available for star formation, a galaxy will
gradually self-enrich as the newly produced stars pollute the
interstellar medium with metals through mass loss and supernovae
winds. In this simplistic assumption the spread of a metallicity distribution for a single population should be related to the time taken to produce that population. Consequently, the `gap' between peak metallicities of the two populations should be related to the difference in their formation times. 

In the left panel of Fig.~\ref{enrichment1} we show the separation
between the metal-rich and the metal-poor population peaks,
$\Delta\mathrm{[Fe/H]}$, as a function of the approximate time when the
metal-poor stars stopped forming, defined as the 95$^{\mathrm{th}}$
percentile of the metal-poor stellar particle formation lookback
times, $\mathrm{t_{mp,95}}$. The points are coloured by the fraction
of metal-poor stars within the galaxy. The left panel displays the field galaxies and the right panel shows the satellite dwarfs. In case of the field galaxies a very clear trend is seen,
where higher metallicity separations occur in galaxies that form their metal-poor population quickly, thus allowing the metal-rich
population to build up whilst star formation is ongoing. This trend is also seen for satellite dwarfs, 
but with significant scatter. It appears that the metal-poor
population in satellites forms later than in field dwarfs, typically
around $\sim$7 Gyr in lookback time. However, for the same metal-poor
population formation time we now see a wide range of fractions of
metal-poor stars. The mechanism responsible for the formation of the two
populations in these galaxies must then also cause star formation to
stall so that only relatively few metal-rich stars are able to form. We have established that the infall of the satellite into the host halo and the stripping of its available gas is capable of providing these conditions.

The triangles mark the galaxies which are spatially segregated. It is evident that spatial
segregation may occur for any metallicity peak separation. However, in
the limit where the metal-poor population is strongly dominant by mass
or the metal-rich stars have little time to form, one might also
expect to see strong spatial segregation due to the first metal-rich stars preferentially forming in central high-density regions. This is evident in the case of satellite galaxies, where the majority of galaxies with two well-segregated populations have very large fractions of
metal-poor stars and are clearly distinct from the rest of the
sample. 

\subsection{The transition between the metal-poor and the metal-rich population}
\label{dip}
\begin{figure}
		\centering
		\includegraphics[width=\columnwidth]{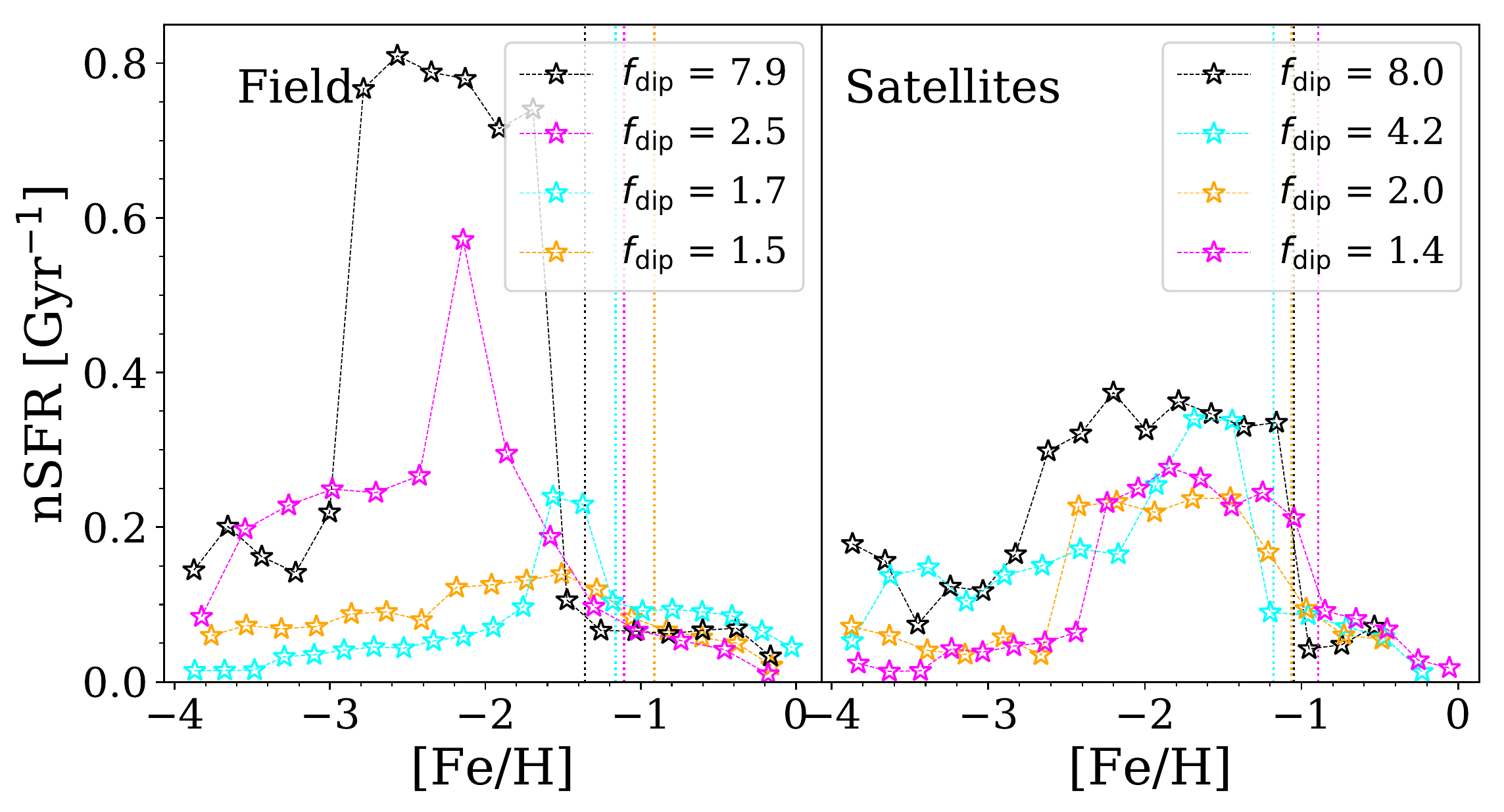}

		\caption{The median star
                  formation rate in bins of stellar metallicity,
                  normalised by the stellar mass of the galaxy, shown for
                  four example galaxies with metal-poor peak to `dip'
                  ratio, $f_\mathrm{dip}$, given in the
                  legend. The values represent how quickly the stars in each metallicity bin form. }

		\label{enrichment2}
\end{figure}

What causes the `dip' in the metallicity distribution that allows us
to define a boundary between two populations? 

In Fig.~\ref{enrichment2} we investigate four
field and four satellite galaxies with well defined `dips'. We define $f_{\mathrm{dip}}$ as the
ratio of the height of the metal-poor population peak to the height of
the distribution at which the metallicity cut is placed in a normalised [Fe/H] histogram. We show the
median star formation rate normalised by the total stellar mass, $\mathrm{nSFR}$,
for bins in metallicity. The dotted vertical lines show the
metallicity at which we place the cut between the two populations. It
is evident that larger `dips' correspond to larger drops in star
formation rate near the boundary between the two populations. The
metal-poor stars appear to have formed at a larger than average rate,
while for the metal-rich population formation, the star formation rate
drops significantly. Note that after the drop in star formation the stars form
at an approximately constant rate for field dwarfs, such that the metal-rich population
takes a long time to complete its formation, which is consistent with
the origin of the metallicty `gap' proposed earlier. For the satellite galaxies, one can see
a gradual decrease in the star formation rate following the
drop. This gradual decrease in star formation activity is consistent with rapid removal of gas following infall in satellites.
\citep{SimpsonQuench}.

\subsection{Where is the gas enriched?}
\label{enrichgas}

\begin{figure}
		\centering
		\includegraphics[width=\columnwidth]{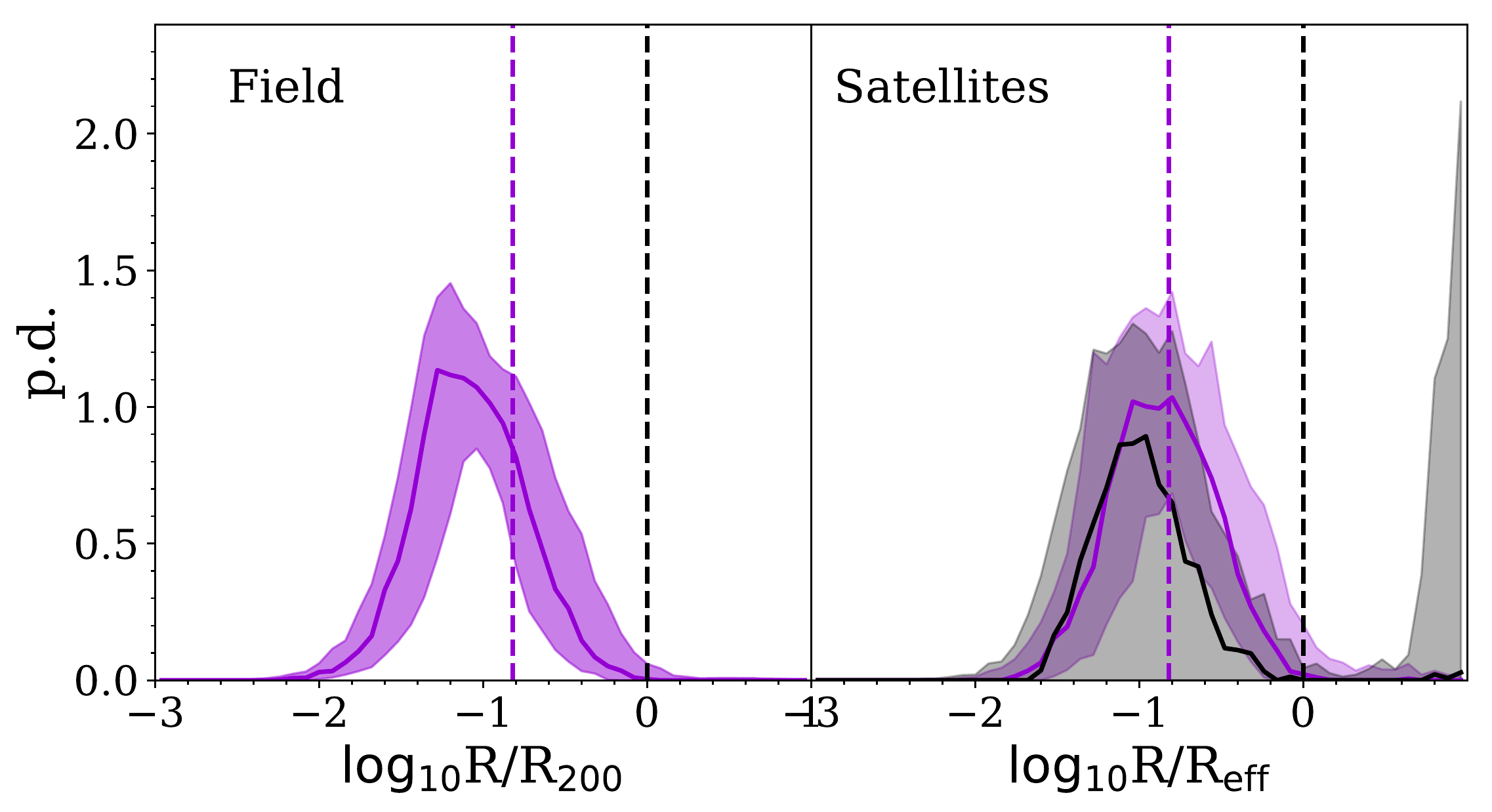}

		\caption{The position of the gas
                  particles destined to form a metal-rich star at the
                  time when they crossed the metal-rich metallicity
                  threshold, normalised by the virial radius of the
                  halo. The median histogram shape is shown with a
                  purple line and the bands represent the
                  16$^{\mathrm{th}}$ and the 84$^{\mathrm{th}}$
                  percentiles. The black and purple dashed lines
                  represent the virial radius and the typical galaxy stellar halo
                  size (0.15$\times$R$_{200}$), respectively. }

		\label{enrichment3}
\end{figure}
One may now ask whether it is gas enriched within the galaxy or
elsewhere that causes the formation of the metal-rich stars. In order
to investigate this, we track all gas particles that have spawned a
stellar particle and the evolution of their metallicity. We find the
position of these particles when they first cross the metallicity
threshold that would allow them to be classified as metal-rich
particles. In the left panel of Fig.~\ref{enrichment3}, for field galaxies, we show a
stacked histogram of these positions, together with the
16$^{\mathrm{th}}$ and the 84$^{\mathrm{th}}$ percentiles, normalised
by the virial radius of the halo at that time. The black dashed line shows the location of the virial radius. It can be seen that almost all particles reach the metallicity required to form a metal-rich star within the virial radius of the halo and only $\sim$~1.5 per cent are enriched outside the halo. In a merger each of
the merger partners can carry on forming stars, until the stellar components have also merged into a single stellar halo. The purple dashed line shows the typical galaxy stellar size, 0.15$\times$~R$_{200}$. Approximately 65.5 per cent of gas particles are enriched within the stellar halo of the galaxy and only 33 per cent are enriched outside and within the virial radius. We
thus conclude that the enrichment necessary to form a metal-rich
population is mostly self-enrichment within the galaxy whether it is
in the process of merging or not.

In the right panel of Fig.~\ref{enrichment3}, we show the corresponding distributions for satellite dwarfs. The positions
are normalised by a radius, R$_{\mathrm{eff}}$, which is the virial
radius pre-infall (purple stacked histogram) or the tidal radius of
the satellite after infall (black stacked
histogram). As in the case of field dwarfs, it is clear that prior to infall enrichment occurs within the galaxy, with only $\sim$3 per cent
of the gas being enriched elsewhere. After infall, the individual
histories for each satellite vary significantly, with some
undergoing no further enrichment at all and others showing evidence of
further self-enrichment, as displayed by the grey shaded region. A significant fraction of galaxies appear to
accrete some of their gas after infall (see the 84$^{\mathrm{th}}$ percentile peak outside of the virial radius). A satellite may be able to accrete some of its host's
enriched gas if it is moving at a sufficiently low relative
velocity. We conclude that the formation of the metal-rich particles in satellites
is triggered primarily by star formation activity and by
self-enrichment of gas within the satellite itself.

\section{The gas content of simulated dwarfs}
\label{AppendixB}

\begin{figure}
		\centering
		\includegraphics[width=0.9\columnwidth]{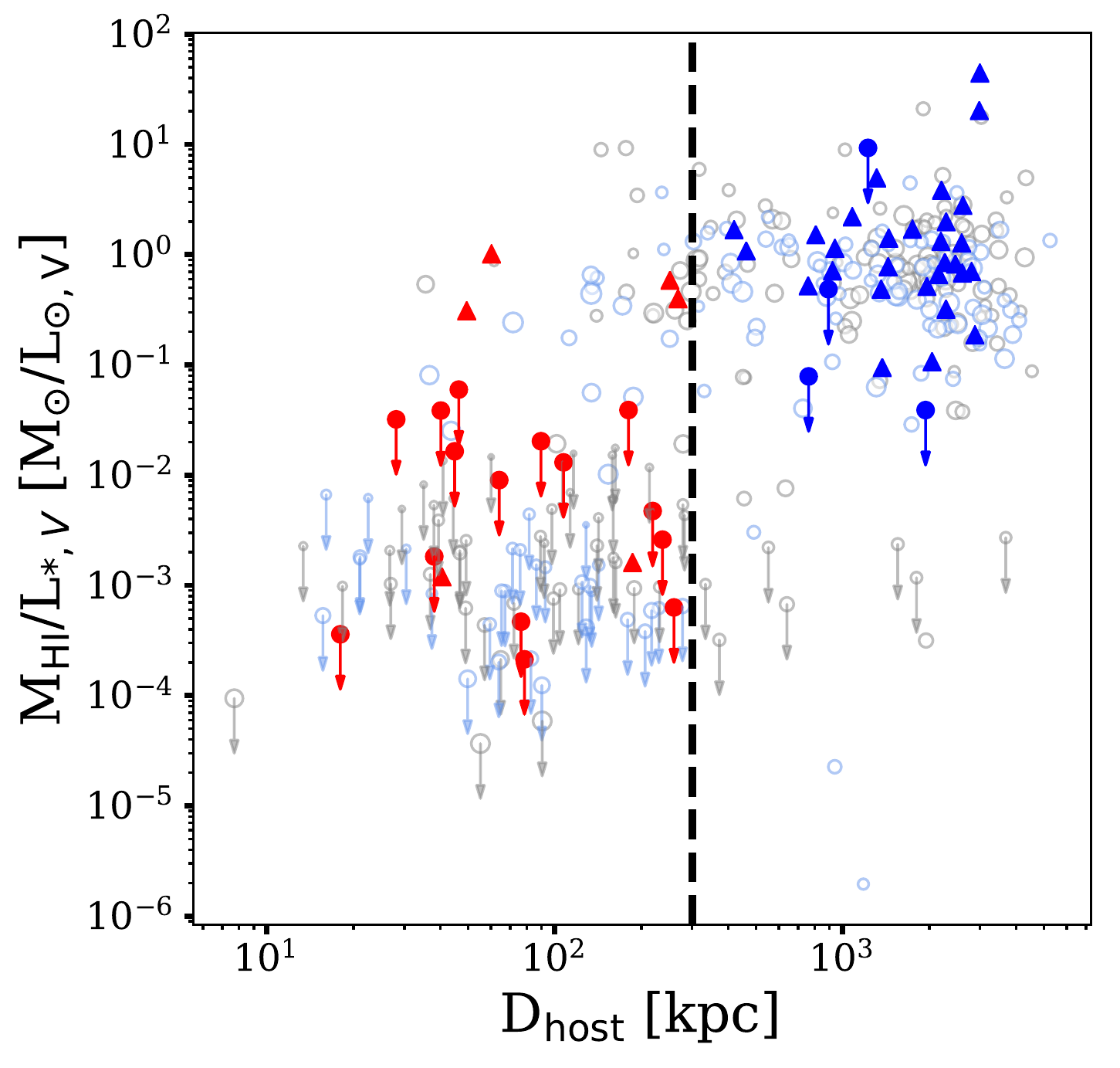}

		\caption{The ratio of mass in H\textsc{i} to the stellar $V$-band luminosity as a function of the dwarf's distance from the nearest host galaxy at $z$~=~0. The grey empty circles are single population galaxies and the blue are two-population dwarfs. The size of the circles is reflective of the logarithm of their stellar mass. The empty circles with arrows represent galaxies with no resolved H\textsc{i} content. Their value is set by the SPH particle mass. The filled red and blue symbols show the measurements for satellites and Local Volume dwarfs, respectively, taken from \citet{mcconnachieCensus} and \citet{spekkens}. Filled symbols with arrows show upper limits. The black dashed line at 300~kpc represents a typical value of the Milky Way's virial radius. }

		\label{morphdens}
\end{figure}

The mechanisms of formation of the two metallicity populations that we identify in this work are largely dependent on the availability of star-forming gas within the dwarfs. It is therefore important to establish whether the H\textsc{i} content of dwarf galaxies within the APOSTLE suite resembles that of the real Local Group dwarfs. 

The determination of H\textsc{i} content would ideally involve the inclusion of radiative transfer schemes and the cold dense ISM within the simulation code. Unfortunately, these schemes are computationally expensive and are not included within the APOSTLE suite. We thus obtain the H\textsc{i} content of the APOSTLE dwarfs using the fitting function of \citet{rahmati} that relates the total photoionization rate to the hydrogen number density at $z$~=~0 and is calibrated on radiative transfer simulations. The atomic hydrogen fraction is then found using a scaling relation between a fraction of molecular hydrogen and gas pressure. The results of applying this prescription can be seen in Fig.~\ref{morphdens}. The observations \citep{mcconnachieCensus,spekkens} are shown with solid symbols and the simulated dwarfs with open symbols. The arrows denote upper limits. It may be seen that the fraction of H\textsc{i} in the APOSTLE dwarfs broadly agrees with the observed values at z~=~0. The satellites (galaxies within 300~kpc of the host) contain little or no gas, whilst the field dwarfs are clustered at M$_{\mathrm{HI}}$/L$_{*,V}$ $\sim$1~M$_{\odot}$/L$_{\odot,V}$, compatible  with observations.

\section{Two populations in Sculptor}

A number of studies have collected spectroscopically determined stellar metal abundances for Milky Way dwarfs \citep{tolstoySculptor,helmisculptor,battagliaspectra,walkersculptor,kirbyrelation}. In particular, many measurements are available for Sculptor and Fornax. \citet{kirbyrelation} have published measurements from medium resolution spectroscopy for 376 Sculptor high-probability members and 676 Fornax members, although for Sculptor the measurements extend to $\sim$350~pc (approximately the half-light radius of this dwarf \citet{mcconnachieCensus}). \citet{walkersculptor} published a larger sample for both dwarfs (1088 stars for Fornax and 1152 for Sculptor with membership probabilities greater than 0.99), with Sculptor members extending to $\sim$2~kpc. According to Fig.~\ref{convergence}, the analysis of the metalllicity distributions should include stars up to 2R$_{\mathrm{hl}}$ to cover the extent of both populations. Thus, the sample of \citet{walkersculptor} would be preferred. However, through identifying overlapping stars in the high-resolution spectroscopy data published in \citet{battagliaspectra}, we find that the magnesium and iron indices relate to [Fe/H] with very large scatter. We therefore choose to apply our method to the \citet{kirbyrelation} data, where we only select dwarfs for which at least 200 measurements are available.

\begin{figure}
    
		\centering
		
		\includegraphics[width=0.9
		\columnwidth]{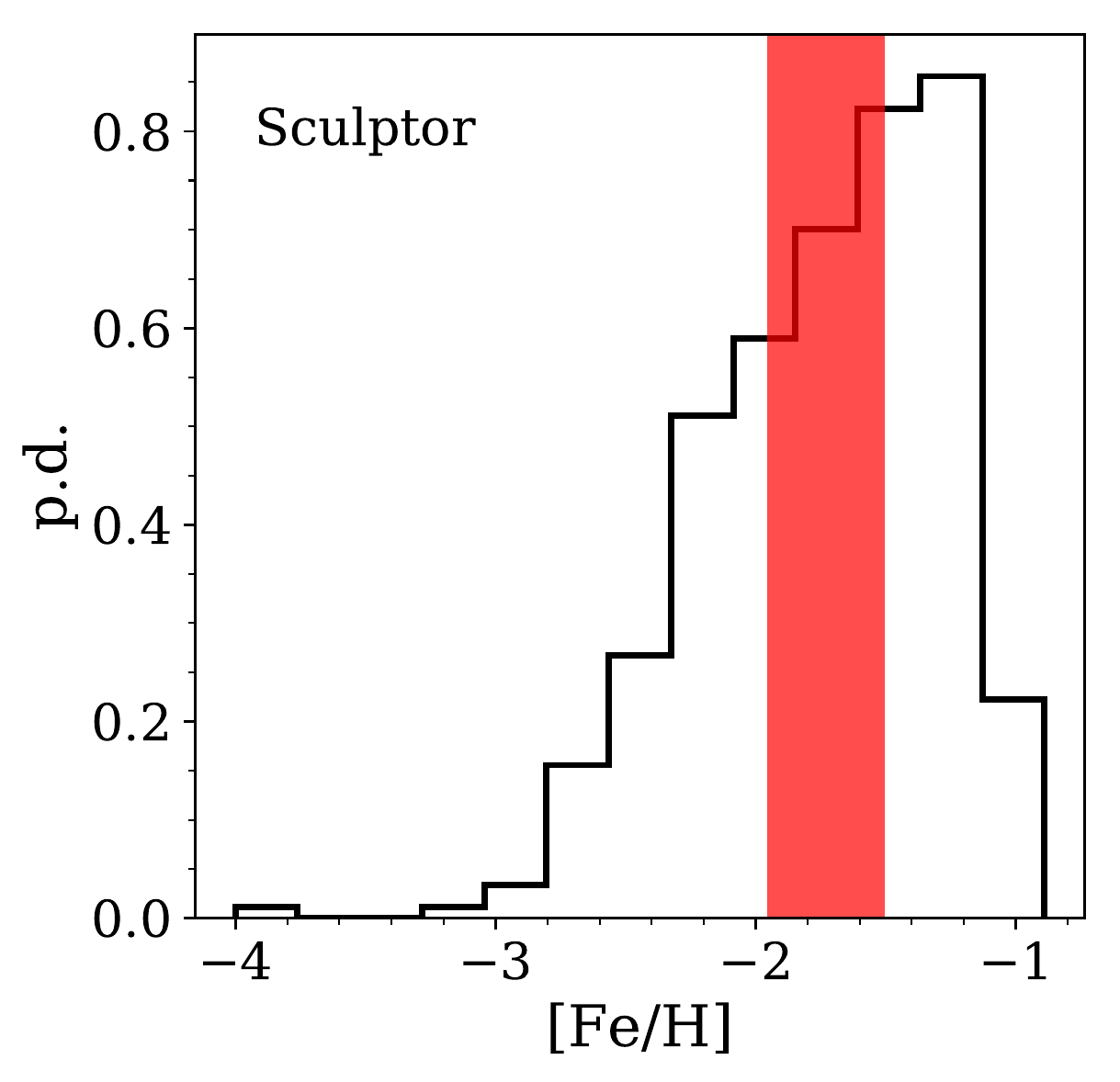}

		\caption{ The metallicity distribution of Sculptor members from \citet{kirbyrelation}. The red band shows the range of cuts, separating the two populations, found with our method.}

		\label{sculptor}
\end{figure}

We sample the errors of each metallicity measurement 1000 times and apply our technique to each sample. We find that none of the dwarfs other than Sculptor display signs of two metallicity populations. For Sculptor, in $\sim$40 per cent of samples two populations can be found. This is shown in Fig.~\ref{sculptor}. Interestingly, the typical cut between the two populations at [Fe/H]$\sim$1.7 (red band) is consistent with that found by \citet{tolstoySculptor} for a more extended sample of stars. Based on the shape of the metallicity distribution alone, with a larger metal-rich population, we lean towards a merger scenario for Sculptor, with consequent gas stripping from the Milky Way halo, preventing the formation of more metal-rich stars.

\label{AppendixC}

\bsp	
\label{lastpage}
\end{document}